\documentclass[pdflatex,sn-nature]{sn-jnl}

\usepackage[version=3]{mhchem} 
\newcommand{\vt}{\vec{\theta}}
\usepackage{xr}
\usepackage{hyperref}
\usepackage{bookmark}

\usepackage{multirow}%
\usepackage{amsmath,amssymb,amsfonts}%
\usepackage{amsthm}%
\usepackage[title]{appendix}%
\usepackage{xcolor}%
\usepackage{textcomp}%
\usepackage{manyfoot}%
\usepackage{booktabs}%
\usepackage{algorithm}%
\usepackage{algorithmicx}%
\usepackage{algpseudocode}%
\usepackage{listings}%

\theoremstyle{thmstyleone}%
\theoremstyle{thmstyletwo}%

\theoremstyle{thmstylethree}%

\begin{document}

\title[Article Title]{Quantum-centric machine learning for  molecular dynamics}

\author[1]{\fnm{Yanxian} \sur{Tao}}
\author[1]{\fnm{Lingyun} \sur{Wan}}
\author*[1]{\fnm{Xiongzhi} \sur{Zeng}}\email{xzzeng@ustc.edu.cn}
\author[1]{\fnm{Yingdi} \sur{Jin}}
\author*[2]{\fnm{Jie} \sur{Liu}}\email{liujie86@ustc.edu.cn}
\author[1,2]{\fnm{Zhenyu} \sur{Li}}
\author*[1,2]{\fnm{Jinlong} \sur{Yang}}\email{jlyang@ustc.edu.cn}

\affil[1]{\orgdiv{State Key Laboratory of Precision and Intelligent Chemistry}, \orgname{University of Science and Technology of China}, \orgaddress{\city{Hefei}, \postcode{230026}, \state{Anhui}, \country{China}}}

\affil[2]{\orgdiv{Hefei National Laboratory}, \orgname{University of Science and Technology of China}, \orgaddress{\city{Hefei}, \postcode{230088}, \country{China}}}



\abstract{Accurate and efficient prediction of electronic wavefunctions is central to {\it ab initio} molecular dynamics (AIMD) and electronic structure theory. However, conventional {\it ab initio} methods require self-consistent optimization of electronic states at every nuclear configuration, leading to prohibitive computational costs, especially for large or strongly correlated systems. Here, we introduce a quantum-centric machine learning (QCML) model—a hybrid quantum–classical framework that integrates parameterized quantum circuits (PQCs) with Transformer-based machine learning to directly predict molecular wavefunctions and quantum observables. By pretraining the Transformer on a diverse dataset of molecules and ansatz types and subsequently fine-tuning it for specific systems, QCML learns transferable mappings between molecular descriptors and PQC parameters, eliminating the need for iterative variational optimization. The pretrained model achieves chemical accuracy in potential energy surfaces, atomic forces, and dipole moments across multiple molecules and ansatzes, and enables efficient AIMD simulations with infrared spectra prediction. This work establishes a scalable and transferable quantum-centric machine learning paradigm, bridging variational quantum algorithms and modern deep learning for next-generation molecular simulation and quantum chemistry applications.}



\maketitle

\section{Introduction}
Accurate prediction of the electronic wavefunction lies at the core of quantum chemistry, particularly in {\it ab initio} molecular dynamics (AIMD), where energies and forces are computed ``on the fly'' from electronic structure calculations. AIMD enables the simulation of chemical processes directly from first principles, without reliance on empirical parameters, thereby allowing detailed investigation of reaction mechanisms, transition states, and other dynamical phenomena. However, each integration step in AIMD requires a self-consistent electronic structure calculation, resulting in prohibitive computational costs for large or complex systems containing hundreds of atoms or requiring long simulation times. The challenge becomes even more severe for strongly correlated systems, where conventional mean-field methods fail and highly accurate electronic structure approaches-such as selected configuration interaction~\cite{HolTubUmr16,LiuHof16,LevHaiTub20,TubFreLev20} \cite{helgaker2013molecular},  and tensor networks \cite{schollwock2005density,verstraete2023density}—are required to properly capture electron correlation effects.

Machine learning (ML) has emerged as a powerful technique to reduce the computational scaling of electronic structure calculations.~\cite{hermann2023ab, doi:10.1021/acs.chemrev.4c00572,D5CS00146C} By efficiently learning high-dimensional mappings between molecular structures and physical observables, ML models can predict properties such as total energy, atomic forces, spectra, atomic charges, ionization potentials and so on \cite{behler2007generalized,bartok2010gaussian,pun2019physically,schutt2017quantum,schutt2018schnet,xu2024hydrocarbon}, with accuracy comparable to high-level quantum chemistry methods but at a fraction of the cost \cite{smith2017ani,zhang2018deep}. Recent advances—including attention mechanisms, message-passing networks, and equivariant neural architectures—have further improved accuracy and generalization in molecular property prediction \cite{kim2017structured,gilmer2017neural,batzner20223}. 
On the other hand, the neural quantum state (NQS) method parameterizes many-body wavefunctions by a neural network, which learns to approximate the complex amplitude of the quantum state from variational principles.\cite{carleo2017solving,pfau2020ab,hermann2023ab,Lange_2024} By optimizing the network parameters to minimize the system’s total energy, NQS achieves variationally accurate representations of ground and excited states.\cite{pfau2020ab,PfaAxeSut24} Despite their remarkable success, NQS still faces significant challenges in accurately representing fermionic antisymmetry and nodal structures, as conventional neural architectures are naturally suited for symmetric functions. Moreover, extending NQS to AIMD presents additional challenges beyond static electronic structure prediction: the need to variationally reoptimize the network parameters along a continuous trajectory of nuclear configurations present a prohibitive computational hurdle. 

Owing to the principles of quantum superposition and entanglement, quantum systems can naturally and efficiently represent many-body wavefunctions that are otherwise intractable classically.~\cite{fauseweh2024quantum,TILLY20221,doi:10.1021/acs.chemrev.8b00803,RevModPhys.92.015003} In digital quantum computing, quantum states can be parameterized using variational ansatzes, such as parameterized quantum circuits (PQCs), where a sequence of gate rotations and entangling operations depends on a set of adjustable parameters\cite{peruzzo2014variational,kandala2017hardware,hempel2018quantum,mcardle2020quantum}. By tuning these parameters, PQCs can approximate molecular ground or excited states, providing a flexible and physically motivated approach to representing quantum systems. One of the major challenges of variational quantum algorithms originates from the high-dimensional optimization problems as the number of ansatz parameters. The combination of quantum computing and machine learning offers a powerful paradigm that leverages the complementary strengths of both fields, where quantum computers naturally encode wavefunctions and machine learning, on the other hand, excels at data fitting.~\cite{TaoZenFan22,YaoJiLi24} By integrating these capabilities, quantum-centric machine learning (QCML) frameworks can learn complex mappings between molecular configurations and quantum states, while exploiting the expressivity of PQCs to represent electronic wavefunctions compactly. 

Like classical ML methods, existing QCML frameworks rely heavily on the availability and quality of training data. Consequently, the ML components must often be retrained when investigating new molecular systems, limiting their efficiency and generalizability. The pretraining–fine-tuning paradigm, widely adopted in large-scale language and vision models, has recently demonstrated remarkable potential in scientific domains.\cite{chithrananda2020chemberta,zhou2023uni,ying2021transformers} By pretraining on large and diverse datasets, a model can acquire broad domain knowledge that can subsequently be refined for specific systems using only a small number of fine-tuning steps. This strategy substantially reduces both data and computational requirements while enhancing transferability and scalability—properties that are essential for practical quantum machine learning (QML) frameworks.

In this work, we pretrain a machine learning model for predicting ansatz parameters based on the Transformer model. The advent of the self-attention mechanism and the Transformer architecture has revolutionized the representational power of ML models. Unlike recurrent or convolutional networks, Transformers capture long-range dependencies without positional constraints, relating any two input or output elements through a constant number of operations.\cite{vaswani2017attention,jiang2024transformer} This architecture exhibits exceptional scalability, parallelizability, and efficiency, and has been successfully applied in quantum chemistry to construct many-body wavefunction ansatzes and to model intricate electron correlations.\cite{viteritti2023transformer,zhang2023transformer,roca2024transformer,shang2025solving} These characteristics make Transformers particularly well suited for learning the parameter manifolds of high-level quantum circuits, enabling transferable and data-efficient QCML models for complex molecular systems.

We employ the Transformer model to learn ansatz parameters across multiple molecules and unitary coupled cluster (UCC) ansatzes within the variational quantum eigensolver (VQE) framework. We demonstrate that the pretrained Transformer model provide an accurate description of the ansatz parameters for equilibrium and dissociated molecular structures. The predicted wavefunctions can then be used to compute total energies, atomic forces, and dipole moments. This Transformer model is successfully adapted to a new molecular system using fewer than 100 fine-tuning epochs with only a small number of training samples, enabling effective generalization with minimal additional computational cost. The high efficiency of the QCML enables us to perform AIMD simulations and the predict infrared spectral based on time-dependent dipole moment trajectories.

\begin{figure}[H]
\centering
\includegraphics[width=1.0\textwidth]{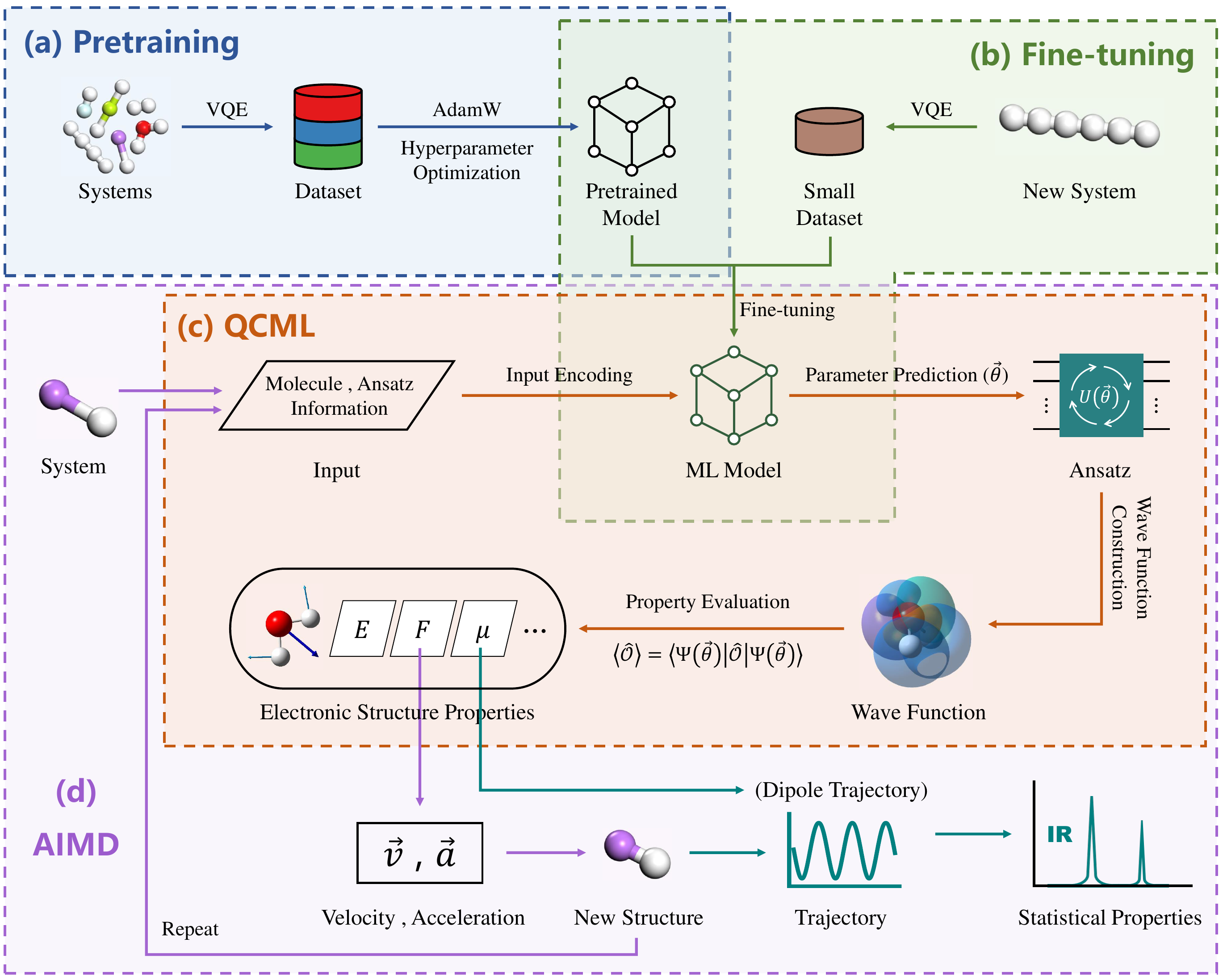}
\caption{Schematic diagram of quantum-centric machine learning (QCML) for molecular dynamics. (a) Pretraining: The variational quantum eigensolver (VQE) algorithm is used to generate the dataset, which is then used to pretrain a Transformer model; (b) Fine-tuning: For new molecules, one can fine-tune the pretrained model using a small dataset, then update the Transformer model that provides an accurate description of ansatz parameters; (c) Property prediction: Given the Transformer model obtained from (b), the electronic wavefunction can be prepared using parameterized quantum circuits and then various molecular properties, such as energies, ionic forces, can be predicted; (d) AIMD: QCML provides highly accurate and efficient evaluations of energies, ionic forces, and dipole moments, enabling long-timescale {\it ab initio} molecular dynamics simulations and reliable spectroscopic predictions.}
\label{fig:toc}
\end{figure}

\section{Results and discussions}

\subsection{Quantum-centric machine learning}
We first introduce the QCML framework for predicting electronic wavefunction and then calculating various electronic structure properties. The overall workflow is shown in Fig.~\ref{fig:toc}. The electronic wavefunction is generally represented as
\begin{equation}
    |\Psi(\vt)\rangle = \hat{U}(\vt) |\Psi_0\rangle,
\end{equation}
where $|\Psi_0\rangle$ denotes the reference state. The unitary transformation $\hat{U}(\vt)$ is in practice mapped onto a parameterized quantum circuit $\hat{U}(\theta_k)\cdots\hat{U}(\theta_1)$ with $\hat{U}(\theta_i)$ being one or two qubit gate. In the standard VQE algorithm, the parameters $\vt$ are variationally optimized by minimizing a certain loss function, e.g. the total energy $E=\langle \Psi(\vt) |\hat{H}| \Psi(\vt)\rangle$, where $\hat{H}$ is the Hamiltonian of a quantum system. This process constitutes a high-dimensional nonlinear optimization problem, where the energy landscape is often non-convex and characterized by multiple local minima.~\cite{lee2018generalized,Arrasmith2021effectofbarren,larocca2025barren} The optimization is further complicated by quantum measurement noise, finite sampling errors, and barren plateaus—regions where the gradient of the cost function vanishes exponentially with system size.~\cite{WanFonCer21} As the number of circuit parameters increases, classical optimizers must navigate a complex and noisy energy surface, leading to slow convergence and substantial computational overhead. These challenges make the nonlinear optimization step the principal bottleneck of VQE, motivating the development of predictive approaches that can infer optimal parameters directly, thereby bypassing or accelerating the iterative variational loop. 

In the QCML, the parameters $\vt$ are predicted using neural networks. The features of the molecular system are encoded into a structured input to the machine learning model $\mathcal{T}$. These include the name of the molecule, internal coordinates (e.g., bond length) used to identify the molecule, and the number of parameters representing the ansatz type. Additional descriptors—such as the number of qubits and gates in the quantum circuit, the number of Pauli strings in the qubit Hamiltonian, the number of electrons, and the energies of frontier molecular orbitals—can be incorporated as auxiliary inputs when necessary.

In previous studies, deep neural network (DNN) models were typically trained for a specific molecule–ansatz pair, requiring independent retraining whenever the molecular system or wavefunction ansatz changed.~\cite{TaoZenFan22,YaoJiLi24} To overcome this limitation and achieve a unified, transferable mapping across diverse chemical systems and ansatz architectures, we introduce a Transformer-based QCML model. Leveraging the self-attention mechanism, the Transformer learns to relate molecular descriptors to the corresponding ansatz parameters in a flexible and data-efficient manner, thereby enabling generalizable prediction of variational quantum circuit parameters across multiple molecules and circuit types. 

We employ the desired output lengths, $N_\theta$, as the weight of the loss of corresponding sample, and define the loss function
\begin{equation}\label{eq:loss}
    \mathrm{Loss} = \frac{1}{N_b} \sum^{N_b}_{k=1}(\frac{1}{S_k} \sum^{S_k}_{j=1}(\frac{1}{N_{\theta^j}} \sum^{N_{\theta^j}}_{i=1}(\theta^j_i-y^j_i)^2) \times w^j),
\end{equation}
where $N_b$ is the number of batches, $S_k$ (set to 64) is the size of $k^{th}$ batch, $N_{\theta^j}$ is the number of $\theta$ (the number of outputs) in the $j^{th}$ sample, $\theta^j_i$ is the value of $i^{th}$ $\theta$ in the $j^{th}$ sample, and $y^j_i$ is the corresponding forecast output, $w^j$ is the weight of the $j^{th}$ sample.

The training data for this model are generated from VQE calculations performed on a diverse set of molecule–ansatz pairs, providing reference variational parameters that encode the intrinsic relationship between molecular structure and quantum circuit configuration (see Methods). The Transformer model is trained by minimizing the loss function of Eq.~\eqref{eq:loss}. In this work, we constructed a relatively comprehensive dataset that includes six molecules with five different UCC ansatzes. In addition, the range of molecular bond length can span the entire molecular dissociation, as shown in Supplementary Table~\ref{tab:dataset}. The inputs of Transformer are those mentioned in the previous section, and outputs are non-zero wavefunction parameters. Because the different ansatz type can already be distinguished by the number of parameters in our dataset, it is not included in inputs. If needed, an ansatz label in text or numeric form can be added to the inputs. The Transformer parameters are trained by the AdamW optimizer,~\cite{loshchilov2019decoupledweightdecayregularization} which is widely adopted in Transformer models due to its stable convergence. 



During the pretraining process, the model is exposed to a broad and diverse dataset encompassing various molecular species, geometries, and ansatz types, allowing it to learn generalizable representations of the underlying structure-parameter relationships. This process enables the model to capture universal patterns in electronic structure and quantum circuit behavior. Subsequently, we consider the fine-tuning process that adapts the pretrained model to a specific molecule or ansatz using only a small subset of additional data. In this stage, the model refines its learned representations to accurately predict ansatz parameters for the target system, achieving high accuracy with minimal computational cost. This hierarchical training strategy effectively eliminates the need to retrain from scratch, reduces the dependence on large quantum datasets, and ensures rapid convergence when extended to new chemical systems or circuit architectures—making it particularly well suited for scalable ab initio molecular dynamics and quantum simulations. A judicious choice of data points that represent distinct physical regions—such as equilibrium, dissociation, and transition configurations—can dramatically improve model compactness and predictive reliability.~\cite{HerStaRiz22,BreHerWes23,Mejuto-Zaera_2023} Such points encode the key changes in electronic structure that define existing phases and their transitions, enabling the fine-tuned machine learning model to reconstruct potential energy surfaces  (PESs) and other electronic structure properties with chemical accuracy across a wide parameter range. This targeted sampling strategy ensures that fine-tuning not only refines the model for a specific system but also enhances its ability to characterize phase behavior and correlation changes in complex molecular and materials systems.

After the machine learning model is trained, the encoded molecular and ansatz information is fed into the Transformer model $\mathcal{T}$, which directly outputs the corresponding variational parameters,
\begin{equation}
\boldsymbol{\theta} = \mathcal{T}(\text{Molecule}, \text{Ansatz}),
\end{equation}
thus eliminating the need for complex high-dimensional optimization on a classical computer. The predicted parameters $\vt$ are then injected into the wavefunction ansatz, i.e., the parameterized quantum circuit $\hat{U}(\vt)$, to generate the molecular wavefunction. By bypassing the iterative optimization loop of the VQE, the QCML achieves orders-of-magnitude acceleration in evaluating molecular properties. Since QCML directly prepare the electronic wavefunction, all relevant quantities—such as total energies, ionic forces, and dipole moments—can be computed efficiently from the resulting state (see Methods). In the following, we demonstrate the capability of QCML by constructing potential energy surfaces, performing AIMD simulations based on atomic forces, evaluating time-dependent dipole moments, and generating infrared spectra. Furthermore, owing to its data-driven foundation, the pretrained Transformer model in QCML readily supports transfer learning to new molecular systems through fine-tuning with only a small number of additional data points, enabling efficient generalization across chemical space.

\subsection{Potential energy surfaces}
\begin{figure}[H]
\centering
\includegraphics[width=1.0\textwidth]{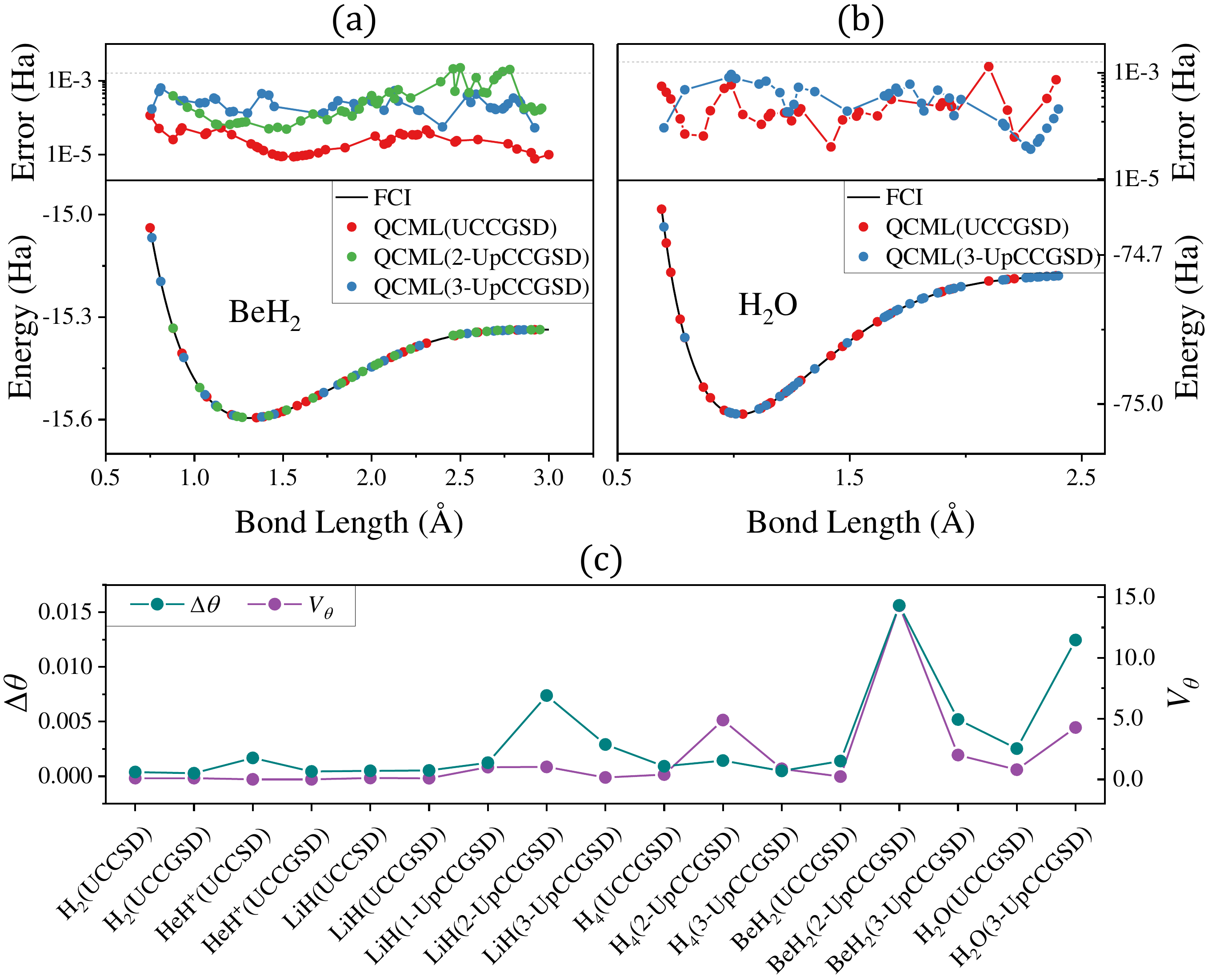}
\caption{Ground-state PESs and their deviation with respect to the FCI results obtained from QCML with different UCC ansatzes for \ce{BeH2} (a) and \ce{H2O} (b). The dotted line indicates chemical accuracy. (c) Comparison between $\Delta \theta$ (the error of the non-zero VQE variational parameters predicted by Transformer model) and $V_{\theta}$ (the sum of the variances of each of the $N_\theta$ non-zero parameters).}
\label{fig:pes}
\end{figure}

We apply QCML to calculate the PESs of small molecules to assess its performance, and the results are shown in Fig.~\ref{fig:pes}, where the exact diagonalization energy, labeled as FCI in the following for simplicity, is taken as reference. It is clear that all PESs predicted by the QCML model are in good agreement with the FCI results, their root mean square deviations (RMSD) are on the order of $10^{-5}$ to $10^{-4}$ Hartree. The discrepancy between QCML and FCI results comes from two parts: One is the intrinsic dataset error, determined by the deviation between VQE and FCI results, which reflects the inherent accuracy of the chosen wavefunction ansatz (see Supplementary Fig.~\ref{fig:err}(a)); Another arises from the prediction error of wavefunction parameters, i.e. the difference between the predicted parameters and the variationally optimized VQE parameters (see Supplementary Fig.~\ref{fig:err}(b)), denoted as the $\Delta \theta$ in Fig.~\ref{fig:pes}(c). These two components affect the fidelity of the constructed quantum state and thus the accuracy of following calculations of electronic structure properties.

To minimize the intrinsic dataset error, we use high-quality training data generated via the VQE using UCC ansatzes. These ansatzes are known to approximate the FCI results with high fidelity for small molecules. Therefore, the error from the dataset itself is negligible, as further supported by the comparison between the error curves in Fig.~\ref{fig:pes} (see Supplementary Fig.~\ref{fig:pes-appendix} for other molecules) and Supplementary Fig.~\ref{fig:err}. On the other hand, the accuracy of the predicted wavefunction depends on several factors: the performance of the Transformer model, the number of non-zero variational parameters $N_\theta$, the selection of training data, and the learnability of the dataset. We optimize the performance of the model through careful hyperparameter tuning as shown in Section~\ref{sec:settings}, and employ a loss function weighted by $N_\theta$ to balance the error of samples with different $N_\theta$ to eliminate the effect of $N_\theta$. Moreover, training data are randomly sampled across bond lengths to ensure good generalization of the model over the full molecular dissociation paths.


After accounting for the error sources discussed above, the remaining discrepancy originates from the learnability of the dataset, which reflects the intrinsic complexity of the function the model must approximate. This learnability can be quantified by the total variance of the non-zero VQE variational parameters across the molecular dissociation range, denoted as $V_{\theta}$ in Fig.~\ref{fig:pes}(c). As shown, systems exhibiting higher $V_{\theta}$ values generally correspond to larger prediction errors, indicating that increased parameter variability imposes a greater challenge for the model to capture the underlying structure–parameter relationship accurately. For example, in the case of the \ce{H2O} molecule, the $V_{\theta}$ for 3-UpCCGSD and UCCGSD ansatzes are 0.1652 and 0.0087, respectively. The $\Delta \theta$ of 3-UpCCGSD is significantly larger than that of UCCGSD, being $1.2 \times 10^{-2}$ compared $2.5 \times 10^{-3}$. Despite these differences in sample complexity across the dataset, QCML consistently achieves $\Delta \theta$ on the order of $10^{-2}-10^{-3}$ for most systems, which is fully sufficient for accurate computation of electronic structure properties such as energy, atomic force, and dipole moment.

\begin{figure}[H]
\centering
\includegraphics[width=1.0\textwidth]{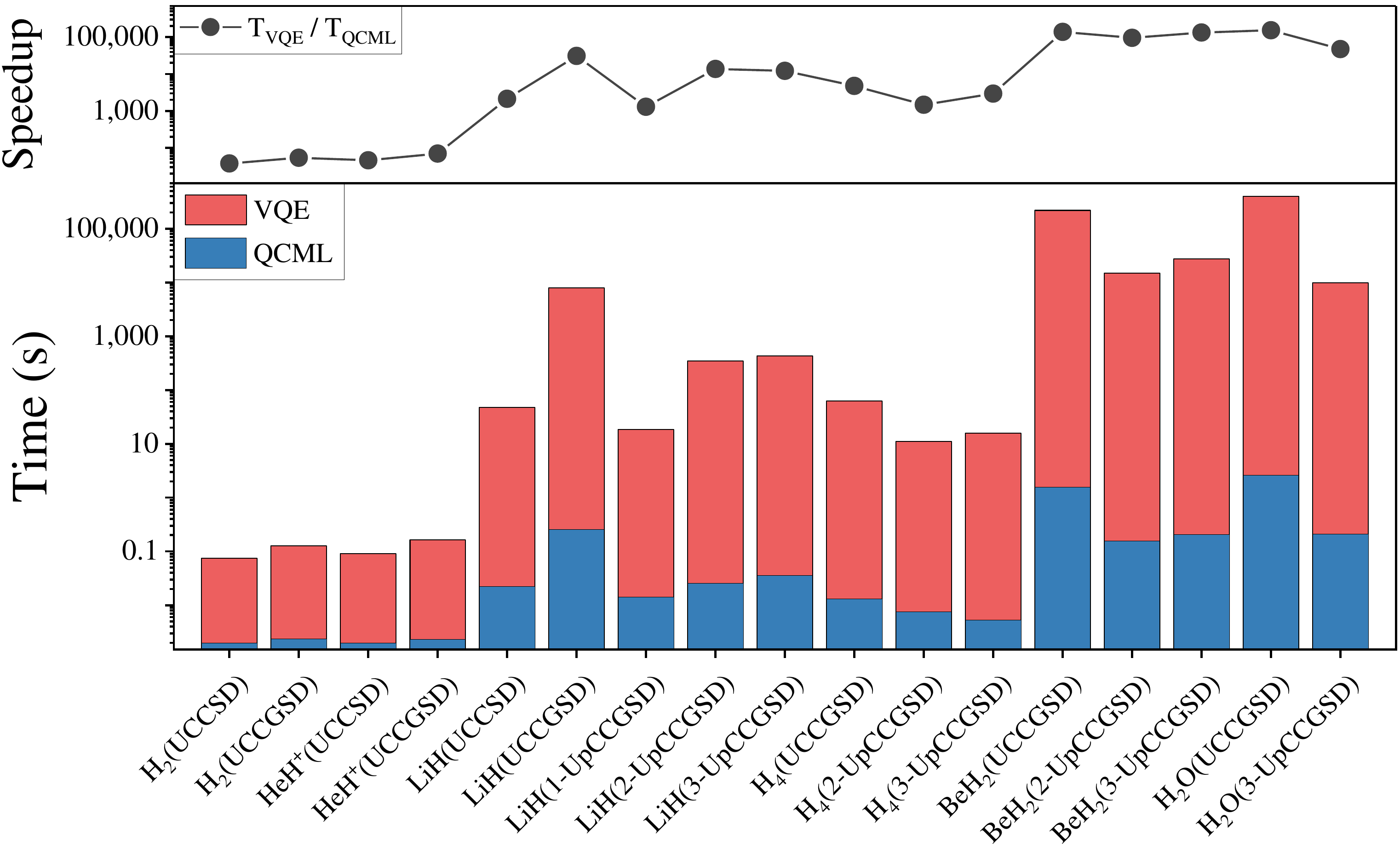}
\caption{Comparison of the average time cost (in second) for computing the single-point energy using VQE and QCML. The black line shows the ratio of the time required for two methods, that is, the speedup of QCML over VQE.}
\label{fig:time-compare}
\end{figure}

The QCML framework eliminates the need for high-dimensional parameter optimization and thereby achieving a drastic reduction in computational cost compared with the standard VQE. Figure~\ref{fig:time-compare} reports the average time required for evaluating a single point energy along molecular dissociation curves using QCML and VQE. Across all tested systems, QCML consistently outperforms VQE by orders of magnitude. For example, the longest time required for QCML occurs for the \ce{H2O} molecule with the UCCGSD ansatz, taking only about 2.6 seconds. In contrast, the corresponding VQE calculation takes 4.5 days, let alone the time required to generate an entire PES. Even in the case with the smallest time difference, that is the \ce{H2} and \ce{HeH+} molecules with UCCSD ansatz, the VQE algorithm still spends 45 times longer than QCML. Training the Transformer model requires approximately 1.5 hours, with the number of epochs set to 3000 to ensure convergence. Notably, the loss function had already plateaued by about 600 epochs, suggesting that the effective training time can be considerably shorter. While model training can be longer than the evaluation of small systems, the trained model is transferable and can be reused without retraining, rendering the training cost negligible in comparison with the dramatic time savings during molecular properties evaluations.

\subsection{Transferability}
\subsubsection{Cross-structural transferability}
\begin{figure}[H]
\centering
\includegraphics[width=1.0\textwidth]{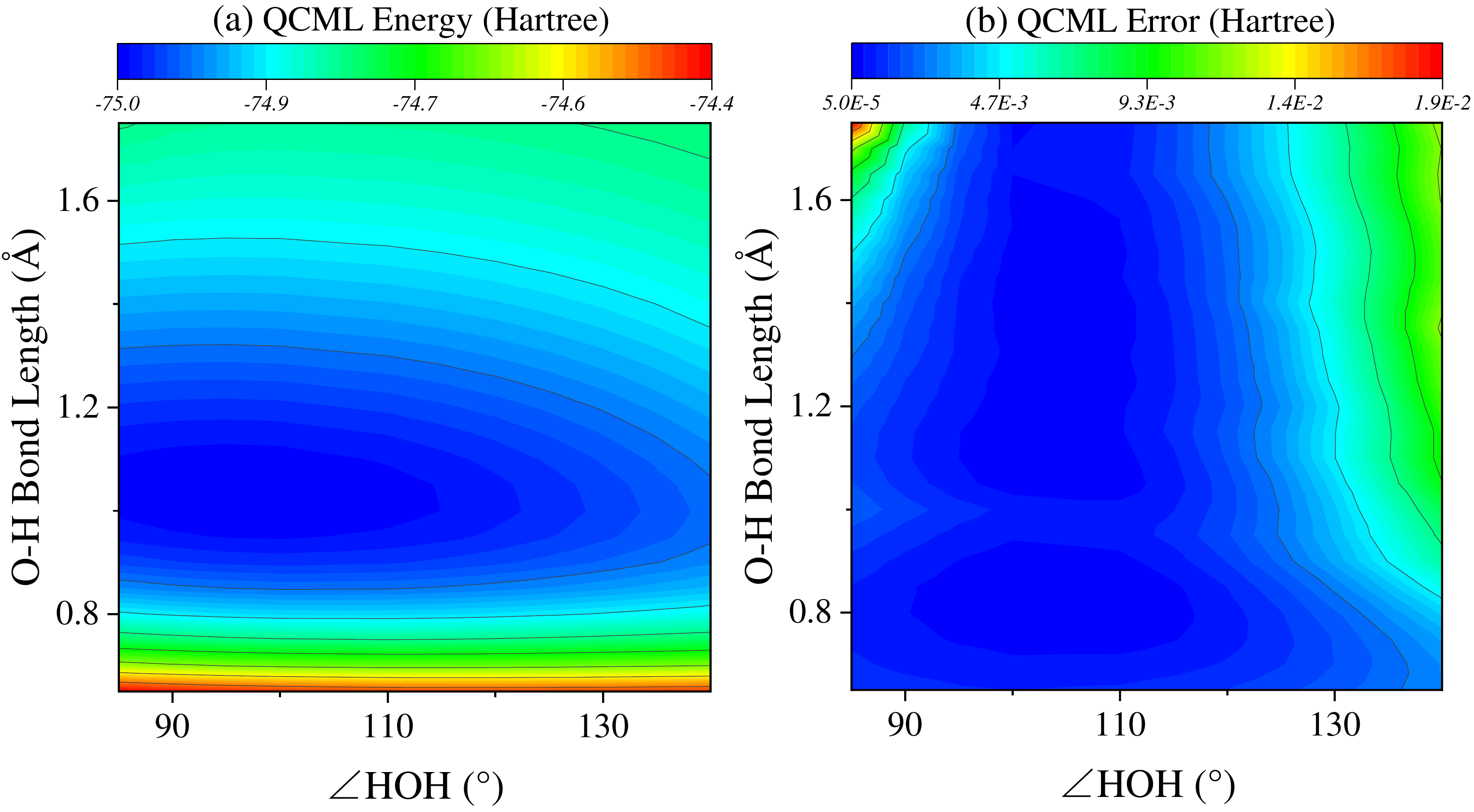}
\caption{Two-dimensional PES as a function of both bond length and bond angle obtained from QCML with a pretrained model for \ce{H2O} molecule, and the corresponding error distribution.}
\label{fig:transfer-h2o}
\end{figure}

We apply the QCML to predict two-dimensional PESs of the water molecule by simultaneously varying the bond length and bond angle. Note that the training set only includes water molecule configurations with varying bond lengths, while the bond angle is fixed at $104.5^\circ$. QCML with a pretrained model can directly generate a full two-dimensional PES as a function of both bond length and bond angle, yielding results that closely match the FCI results, as shown in Fig.~\ref{fig:transfer-h2o}(a), with the corresponding error distribution displayed in Fig.~\ref{fig:transfer-h2o}(b). Although the pretrained model was trained solely on data with a fixed angle of $104.5^\circ$, it generalizes accurately to a broad range of bond angles ($85^\circ - 140^\circ$). In particular, within the typical sampling region near equilibrium geometries relevant to ground-state AIMD simulations, the energy error remains within the $10^{-4}$ Hartree. This demonstrates that the QCML with a pretrained model possesses smooth generalization and strong transferability across molecular geometries. Furthermore, as discussed above, QCML requires only basic molecular and ansatz information, along with a few physical quantities obtainable from Hartree–Fock (HF) calculations, to directly predict wavefunctions and efficiently evaluate the desired electronic structure properties. Consequently, constructing a complete two-dimensional PES incurs only HF-level computational cost while achieving chemical accuracy.

\subsubsection{Cross-system transferability with fine-tuning}
Explicitly applying a pretrained model to predict ansatz parameters for new molecules remains a challenging task, as variations in molecular composition, electronic structure, and circuit architecture can significantly alter the optimal parameter landscape. While pretraining enables the model to learn generalizable representations of the relationship between molecular features and variational parameters, its direct transfer to unseen systems may lead to reduced accuracy due to domain shifts. To overcome this limitation, fine-tuning the pretrained model provides an effective strategy for adapting to new molecular environments. By refining the model parameters using a small, judiciously selected subset of training points, the QCML framework can efficiently capture system-specific correlations and maintain chemical accuracy.

\begin{figure}[H]
\centering
\includegraphics[width=1.0\textwidth]{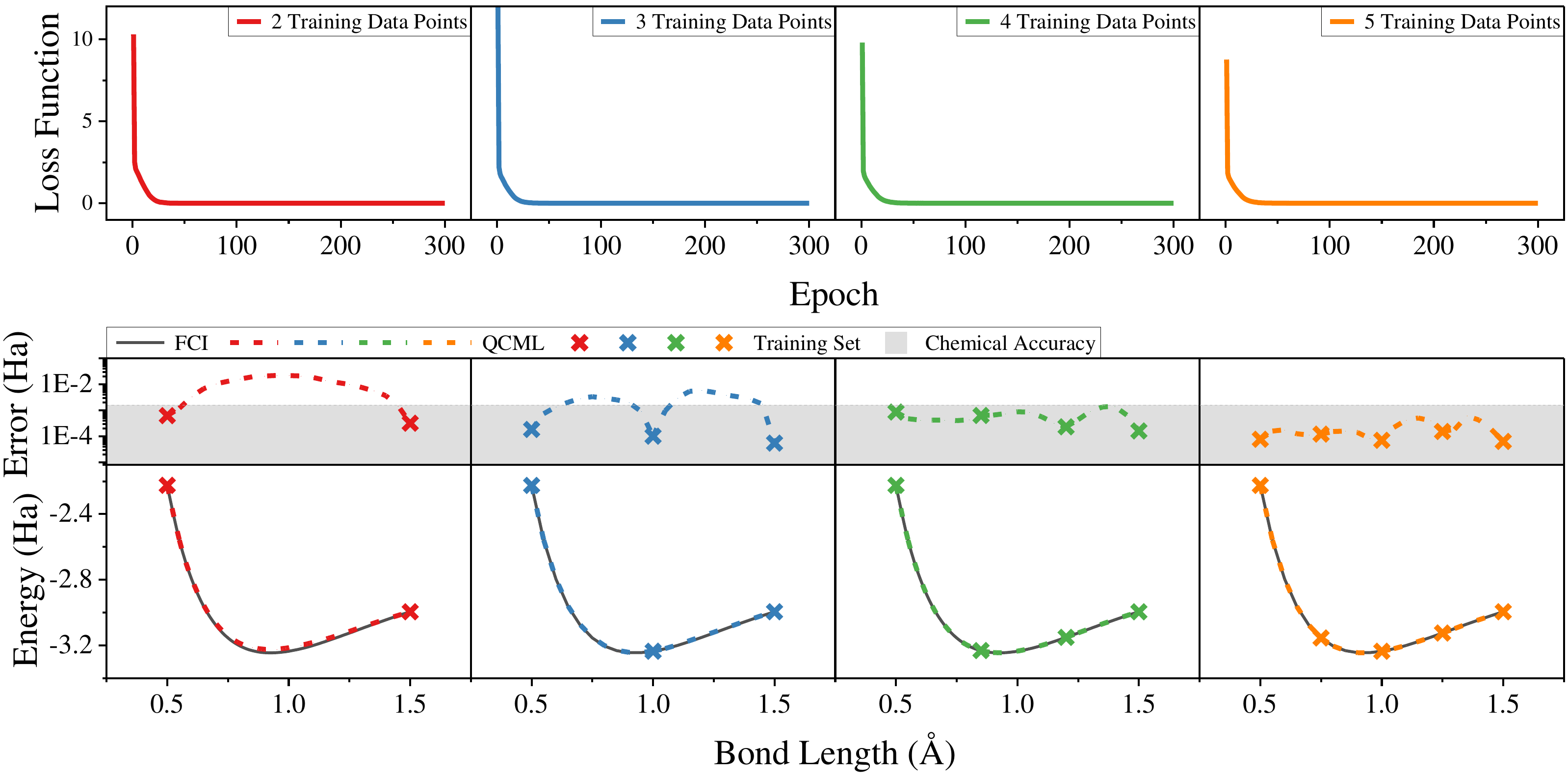}
\caption{Fine-tuning the pretrained model with a small number of data points for the \ce{H6} molecule based on the UCCGSD ansatz. Top panel: Loss function curves during fine-tuning of the pretrained model using different training set sizes (the number of training data points = 2, 3, 4, or 5). Bottom panel: Potential energy surfaces of the \ce{H6} molecule predicted by QCML with the fine-tuned model, along with the error curves between QCML and exact diagonalization results. The colored crosses indicate the training points. The shaded gray region represents the error range within chemical accuracy.}
\label{fig:h6}
\end{figure}

We fine-tune the pretrained QCML model using UCCGSD data for the \ce{H6} molecule and subsequently employ it to generate the PES of this system, as illustrated in Fig.~\ref{fig:h6} and summarized in Table~\ref{tab:h6}. To evaluate the effect of training set size on fine-tuning performance, we systematically vary the number of training data points. When only two to three configurations are included, the model exhibits overfitting and fails to accurately capture the \ce{H6} wavefunction. In contrast, with four training points, QCML reliably predicts the wavefunction and reproduces the PES within chemical accuracy. The RMSD of the predicted energies on the test set is as low as $7.4 \times 10^{-4}$ Hartree, while the wavefunction error remains on the order of $10^{-3}$. As discussed earlier, this level of accuracy is sufficient for computing other electronic-structure properties with confidence. Increasing the number of training samples further improves accuracy, demonstrating smooth convergence. Notably, fine-tuning in QCML is highly data-efficient and computationally lightweight—both training and prediction for \ce{H6} are completed within approximately 10 seconds—highlighting that QCML can be rapidly adapted to new molecular systems at negligible computational cost.

\begin{figure}[H]
\centering
\includegraphics[width=0.8\textwidth]{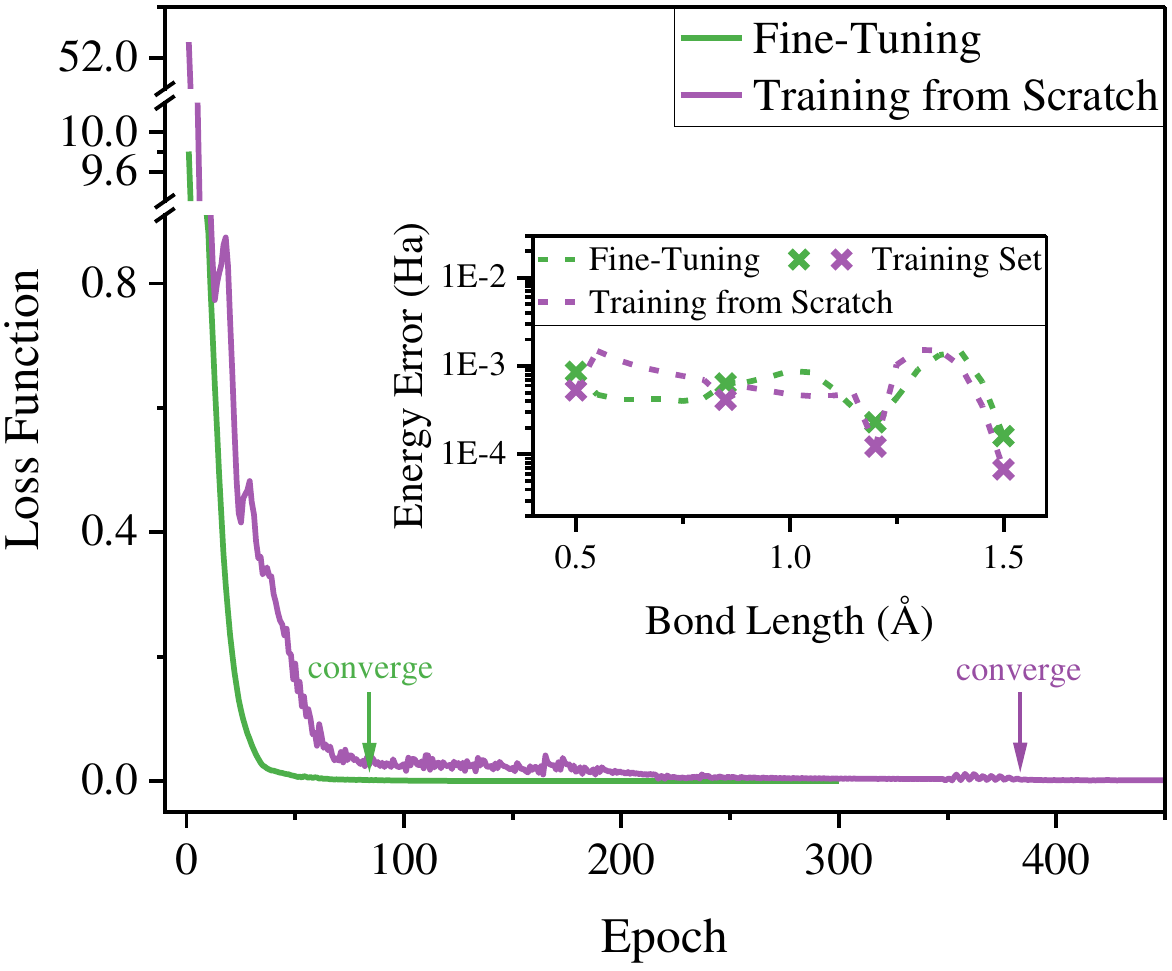}
\caption{The loss function as a function of training epoch for training models from the scratch and pretrained models. The inset represents the energy errors calculated by QCML with different models.}
\label{fig:sctatch vs fine-tuning}
\end{figure}

The advantage of fine-tuning the pretrained model, rather than training QCML from scratch, is illustrated in Fig.~\ref{fig:sctatch vs fine-tuning}. Fine-tuning exhibits a much faster and more stable convergence, as evidenced by the loss function curves: the fine-tuned model converges smoothly within approximately 80 epochs, whereas training from scratch requires about 380 epochs and shows significant intermediate fluctuations, leading to a training time nearly five times longer. This dramatic reduction in convergence time highlights the efficiency of transferring knowledge from the pretrained model. Moreover, fine-tuning not only accelerates optimization but also enhances accuracy and generalization. When trained from scratch, QCML yields an energy RMSD of $3.5 \times 10^{-4}$ Hartree on the training set and $9.3 \times 10^{-4}$ Hartree on the test set, indicating a 2.7-fold discrepancy and clear signs of overfitting. In contrast, the fine-tuned model achieves RMSD values of $5.5 \times 10^{-4}$  Hartree and $7.4 \times 10^{-4}$ Hartree for the training and test sets, respectively—mitigating overfitting while improving overall predictive accuracy. These results demonstrate that fine-tuning not only yields superior performance but also reduces the retraining time by nearly an order of magnitude, making it a practical and scalable approach for extending QCML to new molecular systems.
 
\begin{table}[h]
    \caption{Fine-tuning the pretrained model with different numbers of training data points for the \ce{H6} molecule. $N_\text{point}$ represents the number of training data points. $T_\text{train}$ and $T_\text{predict}$ represent training time and prediction time, respectively. $\Delta \text{E}$ is RMSD between QCML and exact diagonalization energy for test set.}
    \label{tab:h6}
    \begin{tabular}{ccccc}
    \toprule
    $N_\text{point}$ & $T_\text{train}$ (second) & $T_\text{predict}$ (second) & $\Delta \theta$ & $\Delta \text{E}$ (Hartree) \\
    \midrule
    2 & 6.76 & 0.0092 & 3.0E-02 & 1.4E-02\\
    3 & 8.42 & 0.0090 & 1.2E-02 & 3.2E-03\\
    4 & 7.10 & 0.0091 & 8.1E-03 & 7.4E-04\\
    5 & 8.11 & 0.0089 & 6.4E-03 & 3.0E-04\\
    \botrule
    \end{tabular}
\end{table}

\subsection{\textit{Ab initio} molecular dynamics simulations}
AIMD provides a powerful and conceptually rigorous framework for simulating infrared (IR) spectra directly from first principles. AIMD computes interatomic forces and dipole moments ``on the fly'' from electronic structure calculations, ensuring that anharmonic effects, mode coupling, and temperature-dependent phenomena are inherently included. This allows AIMD to capture realistic vibrational dynamics, including overtones, combination bands, and broadening mechanisms that are inaccessible to harmonic or semi-empirical models. Moreover, by evaluating the time-dependent dipole moment along the AIMD trajectory and computing its autocorrelation function, IR spectra can be obtained without explicit normal-mode analysis, making the method applicable to both ordered and disordered systems.

\begin{figure}[H]
\centering
\includegraphics[width=1.0\textwidth]{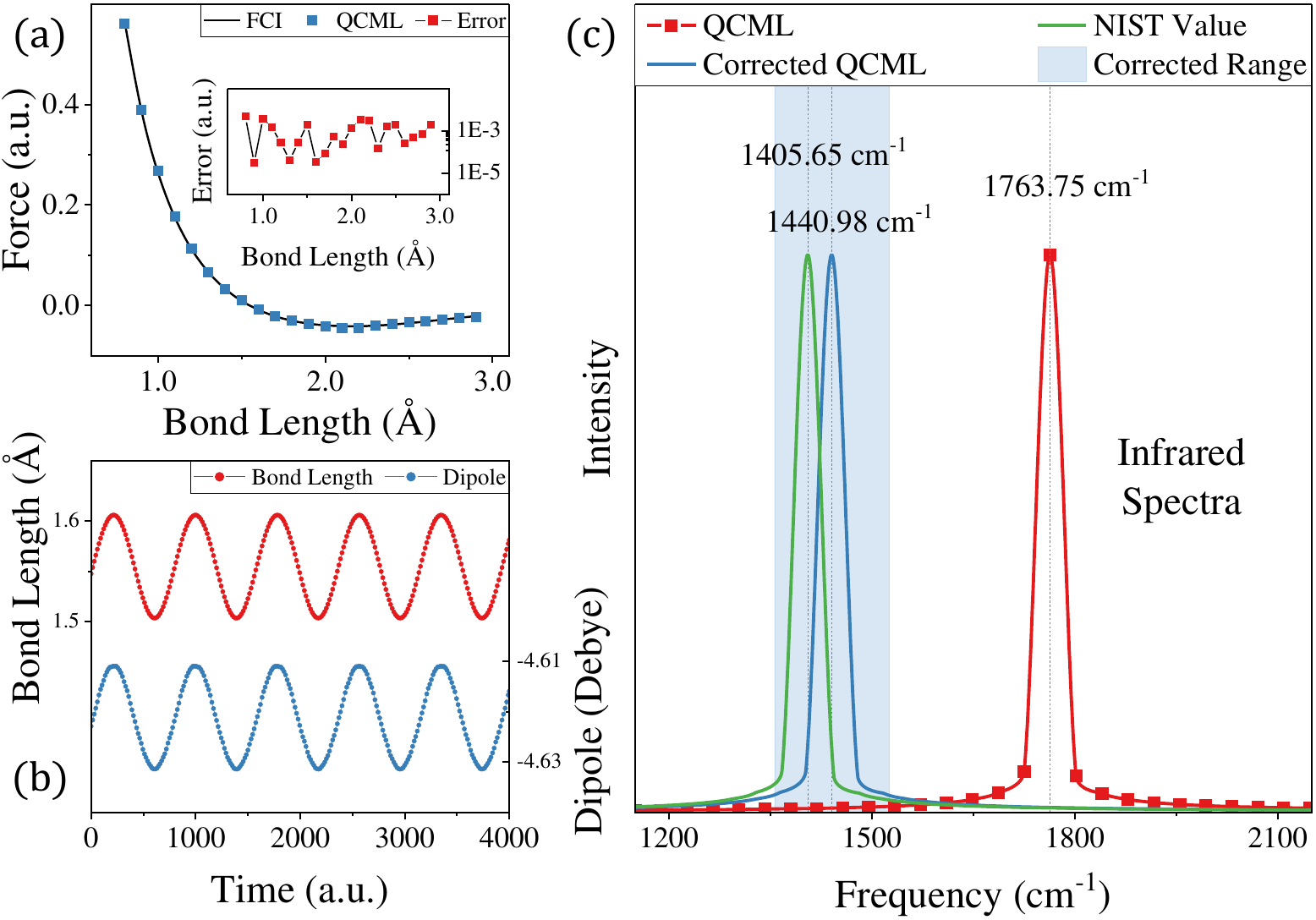}
\caption{(a) The ionic force curve of the \ce{LiH} molecule calculated by QCML. The black curve annotated by FCI is the exact diagonalization results calculated with PySCF software package. (b) Changes in the \ce{LiH} bond length and dipole moment over five vibrational periods during the AIMD simulation. (c) IR spectra of the \ce{LiH} molecule calculated based on QCML. The light blue shading represents the corrected result according to the basis set. And the green line annotated by NIST is the experimental value.}
\label{fig:lihmd}
\end{figure}

Taking the \ce{LiH} molecule as an example, we use QCML to analytically calculate its ionic forces based on the UCCGSD wavefunction ansatz, as shown in Fig.~\ref{fig:lihmd}(a). The inset is the error of the ionic forces with respect to exact diagonalization results. The majority of deviations remain on the order of $10^{-3}$ a.u., a level of accuracy fully sufficient for molecular dynamics simulations. Because QCML can rapidly predict the wavefunction for arbitrary molecular geometries and subsequently evaluate the corresponding electronic structure properties, it can be seamlessly integrated into molecular dynamics simulations to provide continuous analytic forces to ensure both accuracy and stability of dynamical evolution. Figure~\ref{fig:lihmd}(b) exhibits five vibrational periods of AIMD simulations for the \ce{LiH} bond length and dipole moment. The AIMD simulations is performed in the microcanonical (NVE) ensemble, and the initial \ce{Li-H} bond length is set to 1.5475 \AA, which corresponds to the equilibrium bond length of FCI based on STO-3G basis set. The initial velocities of two atoms are determined by the Maxwell–Boltzmann distribution, and are restricted to the Z-direction, that is
\begin{equation}
    v = \sqrt{\frac{3k_B T}{m_{\text{atom}}}},
\end{equation}
where $k_B$ is the Boltzmann constant, $m_{\text{atom}}$ is the atomic mass, and the initial temperature $T$ is set to 300 K. The Velocity Verlet algorithm is employed for the time integration with a time step of 5 a.u. ($\approx 0.12$ fs).

The IR intensity is proportional to the Fourier transformation of the autocorrelation function of the dipole moment, that is
\begin{equation}
    I(\omega) \propto \int \langle \dot{\boldsymbol{\mu}}(\tau) \dot{\boldsymbol{\mu}}(t+\tau) \rangle_{\tau} e^{-i\omega t} dt.
\end{equation}
Therefore, the IR spectra is extracted from a Fourier transformation of the time-dependent dipole moments obtained from QCML, as shown in Fig.~\ref{fig:lihmd}(c). The QCML spectra curve (red) exhibits a prominent absorption peak at the frequency of 1763.75 $cm^{-1}$, corresponding to the \ce{Li-H} stretching vibration. Since the basis set we used is STO-3G, the result can be multiplied by a correction factor $0.817 \pm 0.048$ to obtain the corrected QCML spectra curve (blue) and its confidence region(light blue shading). The absorption frequency after correction is $1440.98 \pm 84.66$ $cm^{-1}$, which is completely consistent with the experimental value 1405.65 $cm^{-1}$ (green)~\cite{10.1063/1.1741916}.

\section{Methods}

\subsection{Theoretical background}

\subsubsection{Variational quantum eigensolver}
The VQE is a hybrid quantum–classical algorithm designed to determine the ground and excited states of a specified many-body Hamiltonian efficiently on near-term quantum hardware.~\cite{peruzzo2014variational,kandala2017hardware,TILLY20221} In VQE, a PQC is used to prepare a trial wavefunction $|\Psi(\vt)\rangle$ where $\vt$ denotes a set of variational parameters controlling gate rotations and entangling operations. The expectation value of the system Hamiltonian, $E(\vt)=\langle \Psi(\vt)|\Psi(\vt)\hat{H}| \rangle$, is evaluated on a quantum device, while a classical optimizer updates $\vt$ iteratively to minimize $E(\vt)$. The second-quantized fermionic Hamiltonian of a quantum system is defined as
\begin{equation}\label{eq:Hamiltonian}
    \hat{H} = \sum_{pq} v^p_q a^\dag_p a_q + \frac{1}{2} \sum_{pqrs} v^{pq}_{sr}  a^\dag_p a^\dag_q a_r a_s.
\end{equation}
One can use Jordan-Winger, or Bravyi-Kitaev transformation to map it into a qubit Hamiltonian
\begin{equation}
    \hat{H} = \sum_{i\alpha} h^i_\alpha \sigma^i_\alpha + \sum_{ij\alpha\beta} h^{ij}_{\alpha \beta} \sigma^i_\alpha \sigma^j_\beta + \cdots.
\end{equation}
In which, $v^p_{q}$ are the one-electron integrals and $v^{pq}_{rs}$ are the two-electron integrals, $a^\dag$ and $a$ are fermionic creation and annihilation operators, respectively. $p,q,r,s$ indicate general spin molecular orbitals, which means that do not distinguish between occupied and unoccupied orbitals. $\sigma$ are Pauli operators and $\alpha, \beta \in \{X,Y,Z,I\}$, $h$ are the corresponding coefficients, $i,j$ denote the index of qubits. 

The VQE is based on the variational principle, which guarantees that the measured energy provides an upper bound to the true ground-state energy. The VQE comprises two key components: the wavefunction ansatz and the optimization algorithm. The choice of ansatz determines the expressivity and physical relevance of the variational space—chemically inspired forms such as the UCC ansatz capture electron correlation systematically, whereas hardware-efficient or problem-tailored circuits aim to balance representational power with circuit depth and noise resilience. In this work, we focus on training the Transformer model with UCC ansatzes, which offer chemically motivated parameterizations with relatively few variational parameters. Note that the QCML framework is not limited to this class of circuits and can be naturally extended to hardware-efficient ansatzes or other variational architectures, providing a flexible foundation for diverse quantum simulation tasks. The second component, classical optimization, iteratively updates the ansatz parameters to minimize the expectation value of the Hamiltonian. The Broyden–Fletcher–Goldfarb–Shanno algorithm is uesd in the VQE calculations that are executed to generate the training data.

\subsubsection{Unitary coupled cluster ansatz}
The UCC ansatz has attracted considerable interest owing to its variational formulation, which makes it naturally compatible with variational quantum algorithms.~\cite{peruzzo2014variational,romero2018strategies,lee2018generalized,anand2022quantum} The UCC wavefunction is wrote as
\begin{equation}
    |\Psi\rangle = e^{\hat{T} - \hat{T}^\dag}|\Psi_0\rangle.
\end{equation}
The widely used UCC singles and doubles (UCCSD) ansatz truncates the cluster operator $\hat{T}$ to the second order
\begin{equation}\label{uccsd}
    \hat{T} = \sum_{ai} t^a_i a^\dagger_a a_i + \frac{1}{4}\sum_{abij} t^{ab}_{ij} a^\dagger_a a^\dagger_b a_i a_j,
\end{equation}
where $i,j$ and $a,b$ indicate occupied and unoccupied spin orbitals, respectively. The coefficients $t$ are to be variationally determined. To further improve accuracy, the UCCGSD ansatz considers generalized singles and doubles operators, namely
\begin{equation}\label{uccgsd}
    \hat{T} = \frac{1}{2}\sum_{pq} t^p_q a^\dagger_p a_q + \frac{1}{4}\sum_{pqrs} t^{pq}_{rs} a^\dagger_p a^\dagger_q a_r a_s.
\end{equation}
The UCCGSD wavefunction is expected to exhibit higher accuracy and robustness than UCCSD for small molecular systems, owing to the explicit inclusion of de-excitation operators that enhance the flexibility of the variational space. ~\cite{Noo00,PieKowFan03,Dav03,Ron03}

In a quantum computing context, the exponential operator is implemented through Trotter-Suzuki decomposition, which maps fermionic excitations to qubit gates. the Trotterized implementation of UCCSD often introduce numerical error with low-order truncation and deep circuit depth. The $k$-fold unitary pair UCCGSD ansatz introduces a pair-coupled-cluster structure combined with generalized excitation operators, and repeats this unitary transformation $k$ times. The pair structure reduces the number of parameters and circuit depth, while the repetition enhances expressivity and systematically recovers correlation effects beyond conventional UCCSD. The wavefunction of $k$-UpCCGSD has a form of
\begin{equation}
    |\Psi\rangle = \prod^k_{l=1} (e^{\hat{T}_1^{(l)}-\hat{T}_1^{(l)\dagger}} e^{\hat{T}_2^{(l)}-\hat{T}_2^{(l)\dagger}}) |\Psi_0\rangle.
\end{equation}
The double excitations here only allow that a pair of electrons are ``excited'' from one spatial orbital to another. Therefore, the cluster operator of $k$-UpCCGSD is
\begin{equation}\label{k-upccgsd}
    \begin{split}
    &\hat{T}_1^{(l)} = \sum_{p q \alpha} t^{p_\alpha,(l)}_{q_\alpha} a^\dagger_{p_\alpha} a_{q_\alpha}, \\
    &\hat{T}_2^{(l)} = \sum_{rs\alpha\beta} t^{r_\alpha r_\beta, (l)}_{s_\alpha s_\beta} a^\dagger_{r_\alpha} a^\dagger_{r_\beta} a_{s_\alpha} a_{s_\beta},
    \end{split}
\end{equation}
where $\alpha$ and $\beta$ indicate the spin of the electron. As such, the cluster operator in the $k$-UpCCGSD are very sparse and its circuit depth scales linearly with the system size. The diversity of available UCC variants—including UCCSD, UCCGSD, and k-UpCCGSD—provides an ideal testbed for validating QCML’s ability to model parameter landscapes across distinct circuit architectures, paving the way for extensions to even more expressive quantum ansatzes. 

As the system size increases, optimizing the parameters of the UCC ansatz becomes a highly nontrivial task due to the enlarged variational space and complex energy landscape. In this work, we initiate the VQE–UCC calculations from the equilibrium molecular geometry, where the Hartree–Fock wavefunction typically provides a reasonable approximation to the exact ground state. Under this condition, the initial amplitudes of the UCC ansatz can be conveniently set to zero. Subsequently, for nearby geometries along the potential energy surface, the optimized parameters obtained from the previous step are used as the initial guess for the next geometry. This sequential initialization strategy provides a reliable initial guess for subsequent VQE calculations, thereby avoiding the need for extensive random sampling of initial parameters.

\subsubsection{Molecular properties}
Once the wavefunction $|\Psi(\vt)\rangle$ is obtained, the expectation value of any observable $\hat{\mathcal{O}}$ can be directly computed as
\begin{equation}
    \langle \hat{\mathcal{O}} \rangle = \langle \Psi(\vt)|\hat{\mathcal{O}}|\Psi(\vt) \rangle.
\end{equation}
The ionic force can be analytically formulated as
\begin{equation}\label{eq:gradient}
\begin{split}
        \frac{\partial E(\vt)}{\partial{R_A}} &= \frac{\partial \langle \Psi(\vt)|\hat{H}|\Psi(\vt) \rangle}{\partial {R_A}}\\
        &=\frac{\partial \langle \Psi(\vt)|}{\partial{R_A}}\hat{H}|\Psi(\vt) \rangle + \langle \Psi(\vt)|\hat{F}_A|\Psi(\vt) \rangle + \langle \Psi(\vt)|\hat{H}\frac{|\partial{\Psi(\vt) \rangle}}{\partial{R_A}}.
\end{split}
\end{equation}
Here,
\begin{equation}\label{eq:force}
    \hat{F}_A = -\sum_{pq} \frac{\partial v^p_q}{\partial R_A} a^\dag_p a_q - \frac{1}{2} \sum_{pqrs} \frac{\partial v^{pq}_{sr}}{\partial R_A} a^\dag_p a^\dag_q a_r a_s-\frac{\partial E_{nuc}}{\partial R_A},
\end{equation}
$A$ labels the atoms, and $R_A$ denotes the coordinate of atom $A$. The $E_{nuc}$ is the nuclear repulsion energy, whose derivative can be easily computed  with $E_{nuc} = -\sum_{A\neq B}\frac{q_1q_2}{|R_A-R_B|}$.
Given the UCCGSD wavefunction is highly accurate for small molecules, the first and third terms in Eq.~\eqref{eq:gradient} can be ignored, that is,
\begin{equation}
    \frac{\partial E(\vt)}{\partial{R_A}} = \langle \Psi(\vt)|\hat{F}_A|\Psi(\vt) \rangle.
\end{equation}
The expression of $\frac{\partial v^p_q}{\partial R_A}$ and $\frac{\partial v^{pq}_{sr}}{\partial R_A}$~\cite{Helgaker1992, 10.1063/1.4927594, doi:10.1021/acs.jctc.2c00440, 10.1063/5.0167444} can be seen in Supplementary Section~\ref{sec:gradint}.

The $\alpha$-components of dipole operator is
\begin{equation}
    \hat{\mu}^\alpha = -\sum_{pq}  \mu_{pq}^\alpha a^\dag_p a_q +\sum_A Q_AR^\alpha_A,
\end{equation}
where $\mu_{pq}^\alpha$ is the dipole matrix in the representation of molecular orbitals with $\alpha\in(x,y,z)$ and $Q_A$ is charge of ion $A$ and $R^\alpha_A$ is $\alpha$-components of $R_A$. The $\mu_{pq}^\alpha$ is defined as
\begin{equation}
    \mu_{pq}^\alpha = \sum_{\mu\nu}C^*_{\mu p} C_{\nu q} \left\langle\mu|\alpha|\nu\right\rangle.
\end{equation}


\subsubsection{Transformer model}\label{sec:settings}
The core principle of the Transformer model is the self-attention mechanism, which dynamically assigns weights to different input elements, capturing global dependencies across the entire input sequence without relying on position-based recurrent structures.
The output of the attention mechanism is computed as a weighted sum of values, where the weights are determined by the compatibility function of the query with key. These weights can be obtained by computing the dot products of the query with all keys
\begin{equation}
    \text{Attention Output} = \text{softmax}(\frac{Q K^T}{\sqrt{d_k}})V,
\end{equation}
where $Q$, $K$ and $V$ represent the query, key, and value matrices, with dimensions $m \times d_k$, $n \times d_k$ and $n \times d_v$, respectively. The greater the compatibility function between the query and key, the larger the weight assigned to the value.




In this work, the Transformer model is implemented using PyTorch.~\cite{NEURIPS2019_bdbca288} The initial learning rate is set to 0.001 and dynamically adjusted according to validation performance using the ReduceLROnPlateau learning rate scheduler. Specifically, when the loss function fails to decrease for 50 consecutive steps, the learning rate is reduced by a factor of two. This adaptive scheduling strategy accelerates model convergence and enhances final training stability. In addition, because the loss function can exhibit significant fluctuations during training, gradient clipping based on the gradient norm is employed to prevent gradient explosion and stabilize optimization. The effectiveness of this stabilization strategy is illustrated in the Supplementary Fig.~\ref{fig:grad_clip}.

\begin{figure}[H]
\centering
\includegraphics[width=1.0\textwidth]{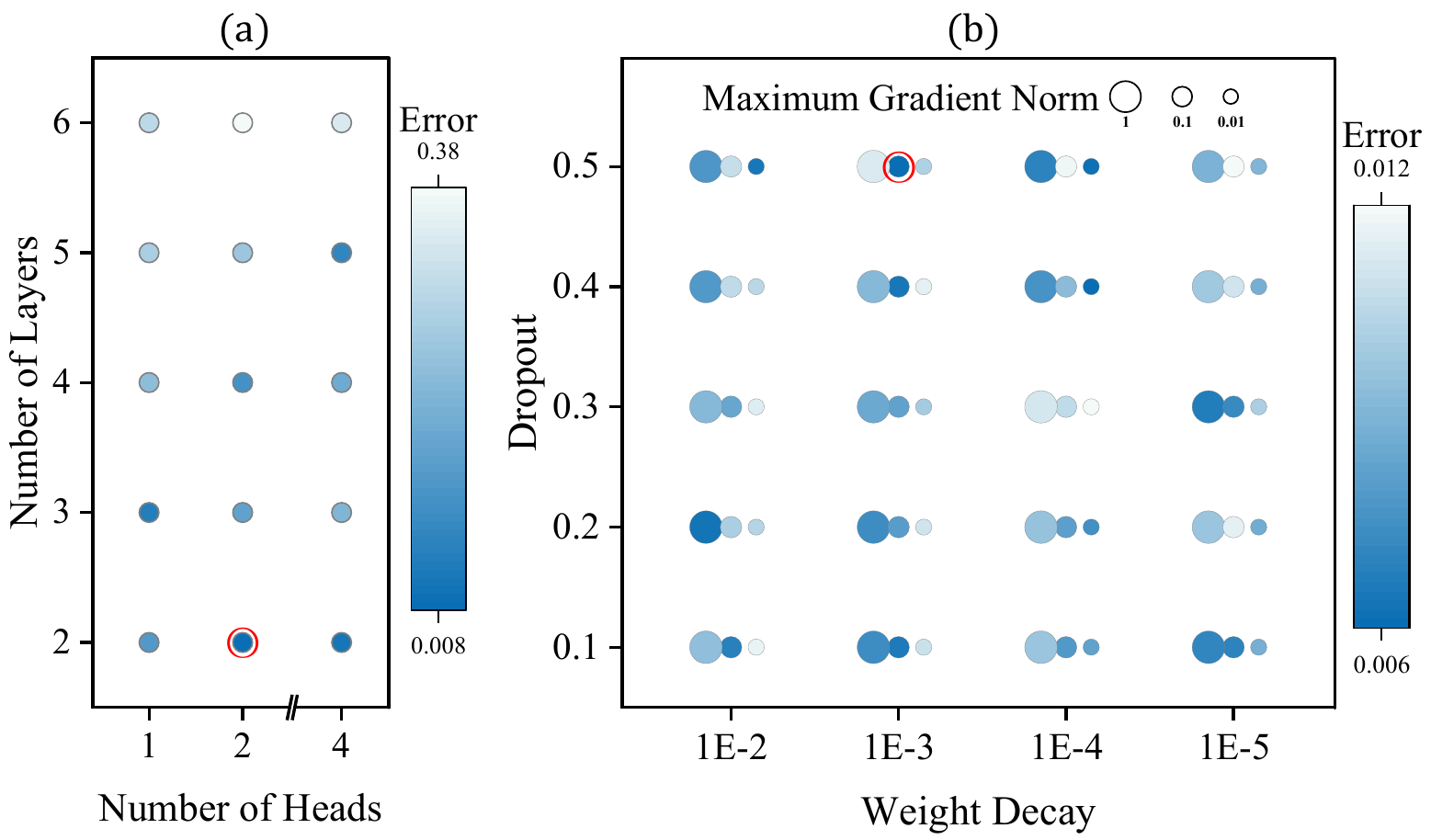}
\caption{Hyperparameter optimization for the Transformer model. The error is represented by the color depth of the point. The darker the point, the smaller the error. The error is RMSD error of non-zero variational parameters predicted by Transformer model. (a) The selection of the number of heads and layers in Transformer model. (b) The selection of dropout rate, weight decay and the maximum gradient norm.}
\label{fig:hyperparams}
\end{figure}

To maximize the performance of the QCML model, we systematically optimize key hyperparameters to identify suitable configurations. Among these, the number of attention heads and Transformer layers primarily determine the model’s size and expressive capacity, and thus exert the greatest influence on performance. As shown in Fig.~\ref{fig:hyperparams}(a), darker points indicate smaller prediction errors, and the point circled in red corresponds to the optimal configuration. The results show that the model with two heads and two layers achieves lower errors than configurations with four heads, while maintaining acceptable computational efficiency. Increasing the number of heads further provides negligible accuracy improvement but substantially increases training time; therefore, both the number of heads and layers are set to two.

Subsequently, we fine-tune additional hyperparameters, including the dropout rate in the Transformer, the weight decay in the AdamW optimizer, and the maximum gradient norm used for gradient clipping, as shown in Fig.~\ref{fig:hyperparams}(b). Since the gradient norms during training remain relatively small, we test maximum values of 1.0, 0.1, and 0.01, and adjust the other parameters within commonly adopted ranges. The optimal hyperparameter set—indicated by the red-circled point in Fig.~\ref{fig:hyperparams}(b)—is selected to balance model accuracy and training stability.

\section{Conclusion}

In summary, we propose a QCML framework that employs a pretrained Transformer model to predict molecular wavefunction ansatz parameters, thereby enabling efficient evaluation of key electronic structure properties such as total energy, ionic forces, and dipole moments. Remarkably, a lightweight Transformer architecture with only two layers and two attention heads is sufficient to accurately predict UCC ansatz parameters across diverse molecular systems. By replacing the classical optimizer in the VQE, our approach eliminates the costly iterative optimization loop and dramatically reduces the number of quantum measurements required, resulting in a substantial reduction in overall computational expense.


Because the number of desired output parameters $N_\theta$ varies substantially among different samples, the $\Delta \theta$ predicted deviations for samples with larger $N_\theta$ tend to be higher than those for smaller ones. To address this imbalance, we define the loss function using $N_\theta$ as a sample-specific weight, ensuring that the normalized parameter deviation $\Delta \theta$ remains comparable across samples. This weighting strategy enhances training stability and mitigates overfitting to systems with fewer variational parameters. The overall error in QCML arises from two primary sources: (i) the intrinsic approximation error of the VQE and (ii) the prediction error of variational parameters produced by the Transformer model. The former can be reduced by employing higher-accuracy ansatzes, such as the UCC family, while the latter can be estimated by the sum of the variances of all predicted parameters $\theta$ within each sample. This decomposition provides a clear basis for quantifying and systematically improving QCML accuracy across diverse molecular systems.

The QCML framework is capable of directly computing PESs or, alternatively, providing high-quality initial guesses for variational parameters in VQE calculations, thereby substantially accelerating subsequent quantum simulations. More importantly, QCML can be seamlessly integrated into molecular dynamics simulations, enabling the analytical evaluation of accurate and continuous atomic forces along nuclear trajectories. By tracking bond-length fluctuations over time, QCML further yields the corresponding time-dependent dipole moments, from which IR spectra can be obtained via Fourier transformation of the dipole trajectories. Together, these capabilities establish a systematic and fully differentiable pipeline from wavefunction prediction to spectroscopic simulation, demonstrating the potential of QCML as a unified, data-driven framework for ab initio molecular modeling and quantum-enabled spectroscopy.

This study demonstrates that fine-tuning the pretrained Transformer model enables QCML to be readily extended to new molecular systems with minimal additional data. Nevertheless, further improving the generalization capability of QCML-so that it can directly predict electronic structure properties of previously unseen systems without retraining-remains a key objective for future work. Notably, excitation amplitudes often exhibit recurring patterns across similar local chemical environments, such as atom pairs, functional groups, and bond types. In addition, hierarchical and nested relationships exist among UCC ansatzes, particularly within the $k$-UpCCGSD family. These insights suggest promising pathways for hierarchical and transfer learning, where models trained on small molecules or local chemical motifs can generalize to larger and more complex systems, and where knowledge gained from lower-order ansatzes can be leveraged to predict higher-order ones. In future research, we aim to further enhance the predictive power of QCML by incorporating more informative molecular descriptors, refining quantum-aware neural architectures, and expanding the training dataset to achieve scalable, transferable, and autonomous quantum simulations across chemical space.

\backmatter

\bibliography{reference}

\section*{Acknowledgements}
This work is supported by Innovation Program for Quantum Science and Technology (2021ZD0303306), the National Natural Science Foundation of China (22073086, 21825302, 22288201,22393913 and 22303090), the Strategic Priority Research Program (XDB0450101) and the robotic AI-Scientist platform of the Chinese Academy of Sciences, Anhui Initiative in Quantum Information Technologies (AHY090400), and the Fundamental Research Funds for the Central Universities (WK2060000018).

\section*{Author Contributions}
J. L. conceived the project of quantum-centric machine learning. Y.T. performed computational work and wrote the initial draft of the manuscript. Y.T., L.W. and X.Z. wrote the code. All authors participated in discussions, contributed ideas throughout the work, and reviewed/edited the manuscript.

\section*{Competing Interests}
The authors declare no competing interests.


\clearpage
\section*{Supplementary Information}







\renewcommand{\tablename}{Supplementary Table}
\renewcommand{\figurename}{Supplementary Fig.}

\setcounter{table}{0}
\setcounter{figure}{0}
\setcounter{section}{0}
\setcounter{equation}{0}
\renewcommand{\theequation}{A\arabic{equation}}








\appendix
\section{Dataset for pretraining}\label{sec:dataset}
The quality of the dataset plays a crucial role in determining the performance of the model. To ensure both accuracy and generalizability, we constructed a comprehensive dataset encompassing six representative molecules - \ce{H2}, \ce{HeH+}, \ce{LiH}, \ce{H4}, \ce{BeH2}, \ce{H2O} - under five widely used ansatzes: UCCSD, UCCGSD, 1-UpCCGSD, 2-UpCCGSD, and 3-UpCCGSD. For each system, the dissociation processes were systematically computed using the VQE algorithm. The scope and size of the pretraining dataset are summarized in Supplementary Table~\ref{tab:dataset}, while the corresponding variations of the VQE variational parameters, i.e., the wavefunction parameters that the pretrained model is designed to learn and predict are shown in Supplementary Fig.~\ref{fig:theta}. This carefully curated dataset provides a solid foundation for achieving both high accuracy and strong transferability in the model training.

\begin{figure}[H]
\centering
\includegraphics[width=1.0\textwidth]{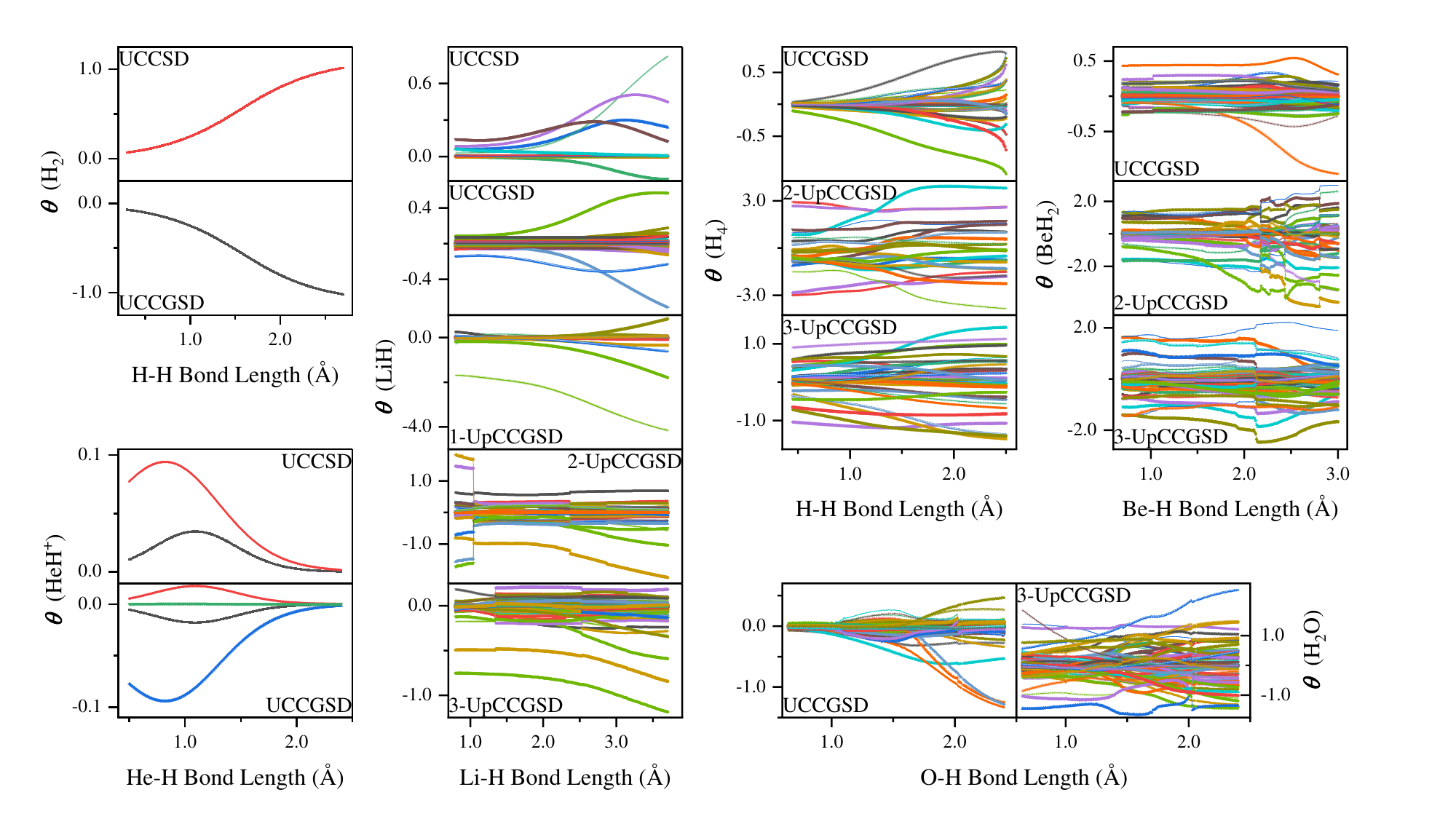}
\caption{The non-zero VQE variational parameters of several molecules under different ansatzes.}
\label{fig:theta}
\end{figure}

\begin{table}[ht]
    \caption{Composition of the dataset. $N_\text{param}$ and $N_\theta$ represent the number of variational parameters and non-zero parameters, respectively. The $V_{\theta}$ column denotes the sum of the variances of each of the $N_\theta$ non-zero parameters.}
    \label{tab:dataset}
    \begin{tabular*}{\textwidth}{@{\extracolsep\fill}c cccccc}
    \toprule
    Molecule & Ansatz & $N_\text{qubit}$ & $N_\text{param}$ & $N_\theta$ & Bond Length (\AA) & $V_{\theta}$ \\
    \midrule
    \multirow{2}{*}{\ce{H2}} & UCCSD  & 4 & 2 & 1 & \multirow{2}{*}{[0.30, 2.70]} & 0.10 \\
                             & UCCGSD & 4 & 4 & 1 &                               & 0.10 \\
    \midrule
    \multirow{2}{*}{\ce{HeH+}} & UCCSD  & 4 & 2 & 2 & \multirow{2}{*}{[0.50, 2.40]} & 0.0014 \\
                               & UCCGSD & 4 & 4 & 4 &                               & 0.0013\\
    \midrule
    \multirow{5}{*}{\ce{LiH}} & UCCSD  & 12 & 44 & 20 & \multirow{5}{*}{[0.80, 3.70]} & 0.11 \\
                              & UCCGSD & 12 & 330 & 111 &                             & 0.0096 \\
                              & 1-UpCCGSD & 12 & 30 & 30 &                            & 1.01 \\
                              & 2-UpCCGSD & 12 & 60 & 60 &                            & 1.02 \\
                              & 3-UpCCGSD & 12 & 90 & 90 &                            & 0.16 \\
    \midrule
    \multirow{3}{*}{\ce{H4}} & UCCGSD & 8 & 66 & 30 & \multirow{3}{*}{[0.45, 2.50]} & 0.40 \\
                             & 2-UpCCGSD & 8 & 24 & 24 &                            & 4.88 \\
                             & 3-UpCCGSD & 8 & 36 & 36 &                            & 0.86 \\
    \midrule
    \multirow{3}{*}{\ce{BeH2}} & UCCGSD & 14 & 609 & 111 & \multirow{3}{*}{[0.70, 3.00]} & 0.23 \\
                               & 2-UpCCGSD & 14 & 84 & 84 &                              & 14.29 \\
                               & 3-UpCCGSD & 14 & 126 & 126 &                            & 2.00 \\
    \midrule
    \multirow{2}{*}{\ce{H2O}} & UCCGSD & 14 & 609 & 189 & \multirow{2}{*}{[0.65, 2.40]} & 0.80 \\
                              & 3-UpCCGSD & 14 & 126 & 126 &                            & 4.27 \\
    \botrule
    \end{tabular*}
\end{table}

\clearpage
\section{Transformer model parameters}
The choice of hyperparameters also has a significant impact on the performance of the model. To ensure stable and efficient training, we have systematically listed all key hyperparameter settings in Supplementary Table~\ref{tab:transformer model}, including the number of attention heads, number of layers, the types and configurations of the optimizer and learning rate scheduler, as well as other parameters that influence model performance.

\begin{table}[hb]
    \caption{Transformer model parameters and settings. $h$ is the number of heads in each layer. $N$ is the number of layers in the encoder and decoder. $d_{model}$ is the output dimension of all sub-layers in the Transformer model.}
    \label{tab:transformer model}
    \begin{tabular}{cc}
    \toprule
    Parameter & Value \\
    \midrule
    $h$          & 2   \\
    $N$          & 2   \\
    $d_{model}$  & 64  \\
    batch\_size   & 64  \\
    loss         & MSE \\
    optimizer    & AdamW \\
    weight\_decay & 0.001 \\
    LR scheduler     & ReduceLROnPlateau \\
    initial value & 0.001 \\
    patience      & 50 \\
    factor        & 0.5 \\
    max\_grad\_norm & 0.1 \\
    dropout         & 0.5 \\
    test\_size      & 0.2 \\
    \botrule
    \end{tabular}
\end{table}

\clearpage
\section{Gradient clipping during training the model}\label{sec:grad_clip}
To mitigate large fluctuations in the loss function and ensure stable model training, we employed gradient clipping based on the gradient norm. Specifically, when the gradient norm exceeded a predefined threshold, i.e., max\_grad\_norm in Supplementary Table~\ref{tab:transformer model}, it was rescaled to prevent excessively large updates. As shown in Supplementary Fig.~\ref{fig:grad_clip}, the loss function curve with gradient clipping (black) is significantly smoother and more stable than that without clipping (red).

\begin{figure}[H]
\centering
\includegraphics[width=0.8\textwidth]{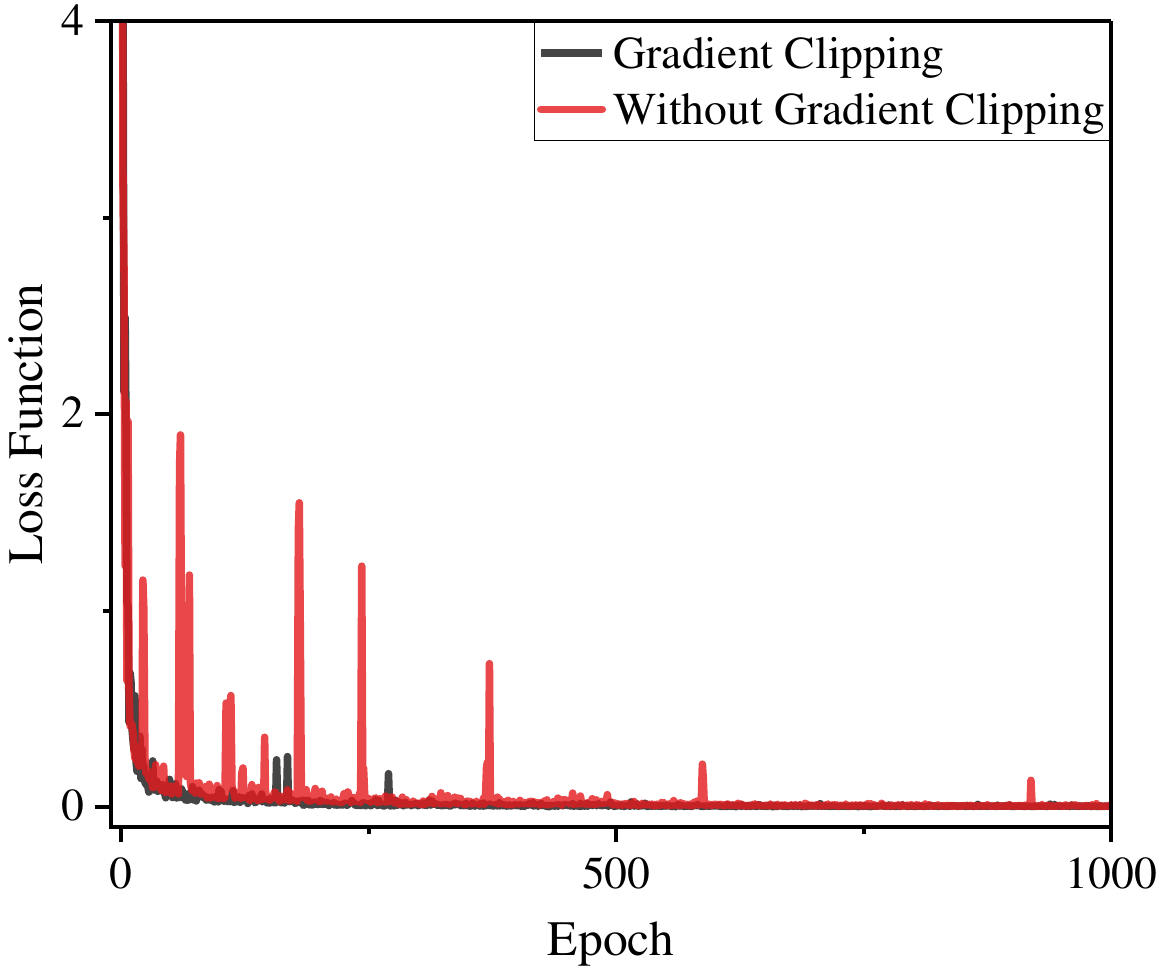}
\caption{The comparison of loss function with and without gradient clipping.}
\label{fig:grad_clip}
\end{figure}

\clearpage
\section{Results of molecular PESs calculations using QCML}\label{sec:pes}
The ground-state PESs of \ce{H2}, \ce{HeH+}, \ce{LiH} and \ce{H4} molecules computed using QCML are shown in Supplementary Fig.~\ref{fig:pes-appendix}. Taking the FCI results as a reference, all deviations remain within chemical accuracy. Supplementary Fig.~\ref{fig:err} illustrates the sources of errors in the QCML PESs. In Supplementary Fig.~\ref{fig:err}(a), the the discrepancies between VQE and FCI reflect the intrinsic errors of the dataset, while in Supplementary Fig.~\ref{fig:err}(b), the differences between QCML and VQE represent the errors arising from the prediction of wavefunction parameters.

\begin{figure}[H]
\centering
\includegraphics[width=1.0\textwidth]{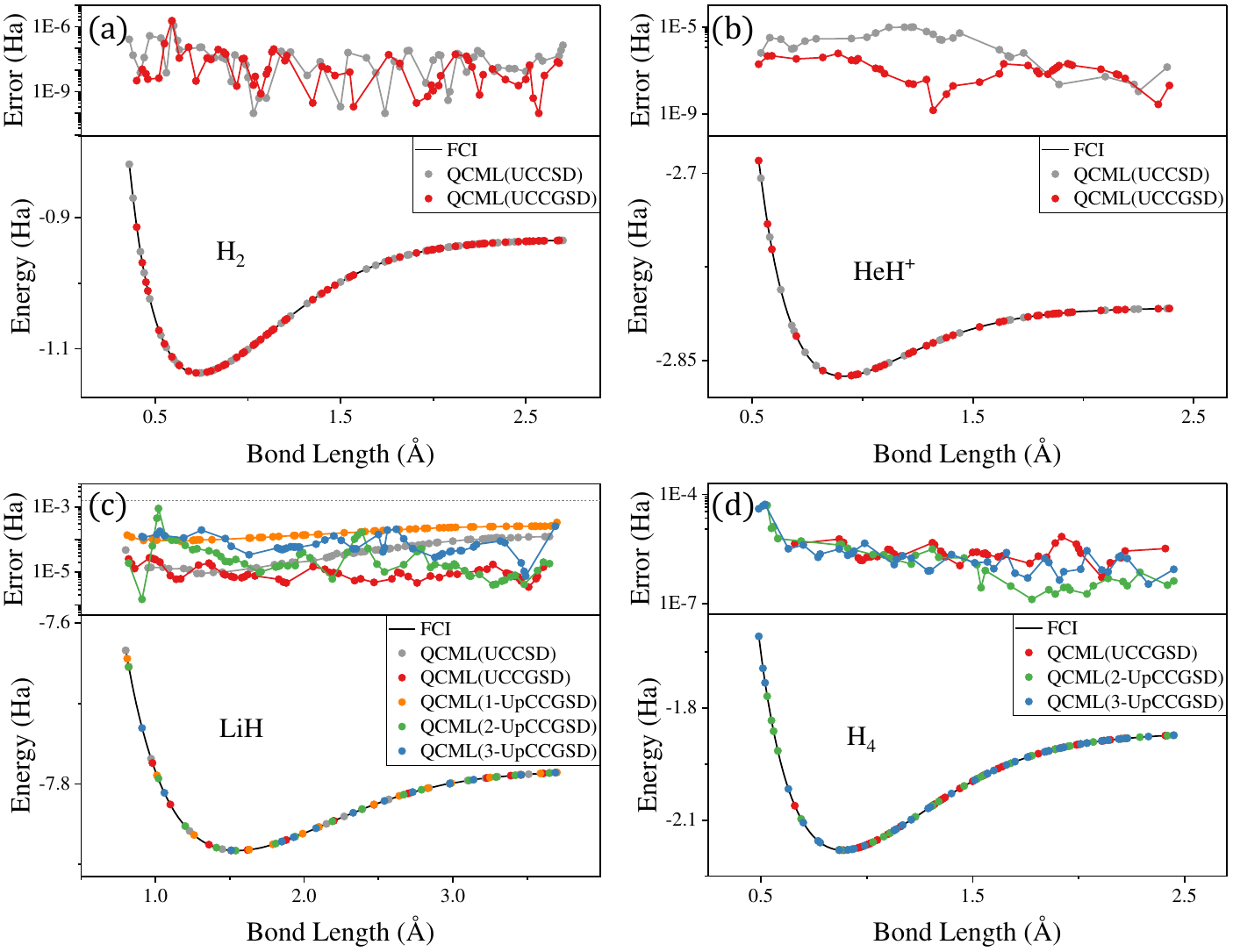}
\caption{Ground-state PESs of \ce{H2} (a), \ce{HeH+} (b), \ce{LiH} (c) and \ce{H4} (d) obtained from QCML with different UCC ansatzes and their difference in energy from the FCI energy. The dotted line in top pictures indicates chemical accuracy.}
\label{fig:pes-appendix}
\end{figure}

\begin{figure}[H]
\centering
\includegraphics[width=1.0\textwidth]{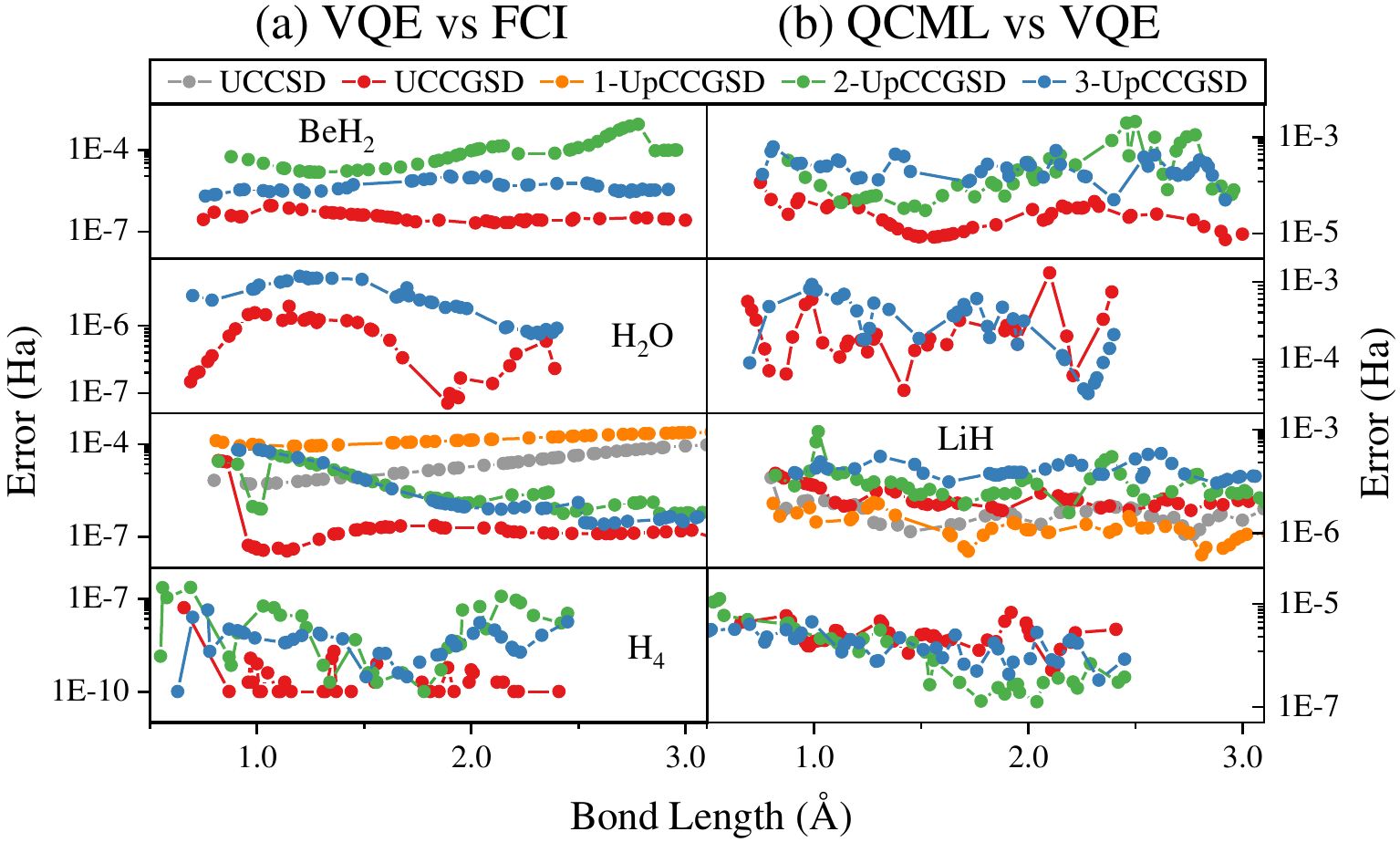}
\caption{The error between the PESs obtained from VQE and FCI (a) and the error between PESs obtained from QCML and VQE (b) for several molecules.}
\label{fig:err}
\end{figure}

\clearpage
\section{Initializing VQE with predicted wavefunction for enhanced efficiency and accuracy}

If higher precision is required, the variational parameters obtained by QCML can be used as the initial parameters in VQE, which can significantly reduce the time cost. For example, when preparing the dataset of \ce{H2O} molecule with 3-UpCCGSD ansatz, we first preform VQE for a part of molecular configurations with initial parameters $\boldsymbol{\theta}_\text{init}=\boldsymbol{0}$ to prepare a small dataset. The time cost is shown as the orange region in Supplementary Fig.~\ref{fig:time}. Then, we train a simple Transformer model on this dataset to predict the VQE variational parameters $\boldsymbol{\theta}_1$ for another part of molecular configurations, and use $\boldsymbol{\theta}_1$ as the initial parameters to preform VQE calculations to reduce the time spent on parameters optimization, as the green region in Supplementary Fig.~\ref{fig:time}. Finally, we repeat above processes to preform VQE calculations for remaining molecular configurations, with the time cost shown as the blue region. As we can see, the average time cost decreases from 5.2 hours ($\boldsymbol{\theta}_\text{init}=\boldsymbol{0}$) to 1.2 hours ($\boldsymbol{\theta}_\text{init}=\boldsymbol{\theta}_2$), greatly reducing the total time required to prepare an entire dataset.

\begin{figure}[H]
\centering
\includegraphics[width=0.8\textwidth]{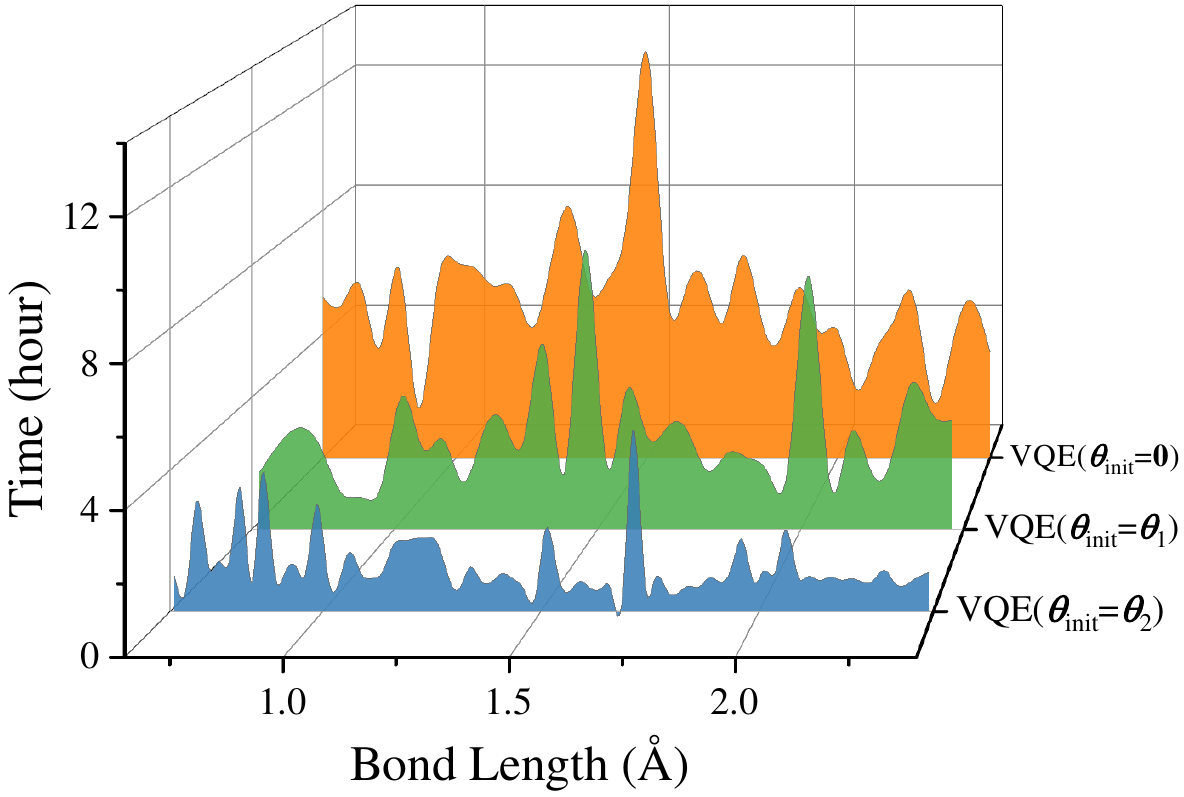}
\caption{Time cost of VQE with different initial parameters.}
\label{fig:time}
\end{figure}

\section{Analytical gradient}\label{sec:gradint}
In this work, the force operator is construct by the first-order analytical derivatives of one-electron and two-electron integrals with respect to nuclear coordinates. The one-electron and two- electron integrals in Eq.~\ref{eq:force} can be expressed in the following form:
\begin{equation}
    v^p_q = \sum_{\mu\nu} C^*_{\mu p}C_{\nu q} h^\mu_\nu,
\end{equation}
\begin{equation}
 v^{pq}_{sr}=\sum_{\mu\nu\sigma\rho}C^*_{\mu p}C^*_{\nu q}C_{\sigma s}C_{\rho r} h^{\mu\nu}_{\sigma\rho},
\end{equation}
where $\mu,\nu,\sigma,\rho$ are used to label the atomic orbitals (AO) and $p,q,r,s$ are used to label the molecular orbitals (MO). Since the integrals $v^p_q$ and $v^{pq}_{sr}$ depend on the electron integrals in the AO basis and the MO coeficients, the first-order derivative of the Hamiltonian operator is in turn correlated with the derivatives of the electron integrals ($h^\mu_\nu$ and $h^{\mu\nu}_{\sigma\rho}$) in the AO basis and the MOs coeficients $C$. The MOs coeficients $C$ have orthogonality:
\begin{equation}
    \sum_{\mu\nu}C^*_{\mu p} S_{\mu\nu} C_{\nu q} = I_{pq}.
\end{equation}
The corresponding derivatives of the electron integrals are calculated by:
\begin{equation}
    \frac{\partial v^p_q}{\partial R_A} = \sum_{\mu\nu} C^*_{\mu p}C_{\nu q} \frac{\partial h^\mu_\nu}{\partial R_A} -\frac{1}{2} \left\{S^{(R)},v \right\}_{pq} ,
\end{equation}
\begin{equation}
    \frac{\partial v^{pq}_{sr}}{\partial R_A} = \sum_{\mu\nu\sigma\rho}C^*_{\mu p}C^*_{\nu q}C_{\sigma s}C_{\rho r} \frac{\partial h^{\mu\nu}_{\sigma\rho}}{\partial R_A} - \frac{1}{2}\left\{S^{(R)},v\right\}_{pqrs}.
\end{equation}
The brace terms are defined as:
\begin{equation}
    \left\{S^{(R)},v \right\}_{pq} = \sum_o \left(S^{(R)}_{po}v^o_q + S^{(R)*}_{qo}v^o_p\right),
\end{equation}
\begin{equation}
    \left\{S^{(R)},v\right\}_{pqrs} = \sum_o \left(S^{(R)}_{po}v^{oq}_{sr} + S^{(R)}_{qo}v^{po}_{sr}+S^{(R)*}_{ro}v^{pq}_{so} + S^{(R)*}_{so}v^{pq}_{or}\right),
\end{equation}
where $S^{(R)}_{pq}$ can be calculated by:
\begin{equation}
    S^{(R)}_{pq} = \sum_{\mu\nu}C^*_{\mu p} C_{\nu q} \frac{\partial S_{\mu\nu}}{\partial R_A}.
\end{equation}




\end{document}